\documentclass[journal]{vgtc}                     

\onlineid{0}

\vgtccategory{Research}

\vgtcpapertype{please specify}

\title{
Perception of Visual Variables on Virtual Wall-Sized Tiled Displays in Immersive Environments  
}

\author{%
    \authororcid{Dongyun Han}{0000-0002-9517-5326},
    \authororcid{Anastasia Bezerianos}{0000-0002-7142-2548},
    \authororcid{Petra Isenberg}{0000-0002-2948-6417},
    and \authororcid{Isaac Cho}{0000-0003-1582-8428}
}

\authorfooter{
  \item
  	Dongyun Han is with Utah State University.
  	E-mail: dongyun.han@usu.edu
  \item
  	Anastasia Bezerianos is with Université Paris-Saclay.
  	E-mail: anastasia.bezerianos@universite-paris-saclay.fr

  \item Petra Isenberg is with Université Paris-Saclay.
  	E-mail: petra.isenberg@inria.fr
    \item
  	Isaac Cho is with Utah State University.
  	E-mail: isaac.cho@usu.edu
}

\abstract{%
We investigate the perception of visual variables on wall-sized tiled displays within an immersive environment. 
We designed and conducted two formal user studies focusing on elementary visualization reading tasks in VR. 
The first study compared three different virtual display arrangements (\studyFactors{Flat, Cylinder,} and \studyFactors{Cockpit}). 
It showed that participants made smaller errors on virtual curved walls (\studyFactors{Cylinder} and \studyFactors{Cockpit}) compared to \studyFactors{Flat}.
Following that, we compared the results with those from a previous study conducted in a real-world setting. 
The comparative analysis showed that virtual curved walls resulted in smaller errors than the real-world flat wall display, but with longer task completion time.
The second study evaluated the impact of four 3D user interaction techniques (\studyFactors{Selection, Walking, Steering,} and \studyFactors{Teleportation}) on performing the elementary task on the virtual \studyFactors{Flat} wall display. The results confirmed that interaction techniques further improved task performance.
 Finally, we discuss the limitations and future work. 
}


\teaser{
  \centering
  \includegraphics[width=1\linewidth]{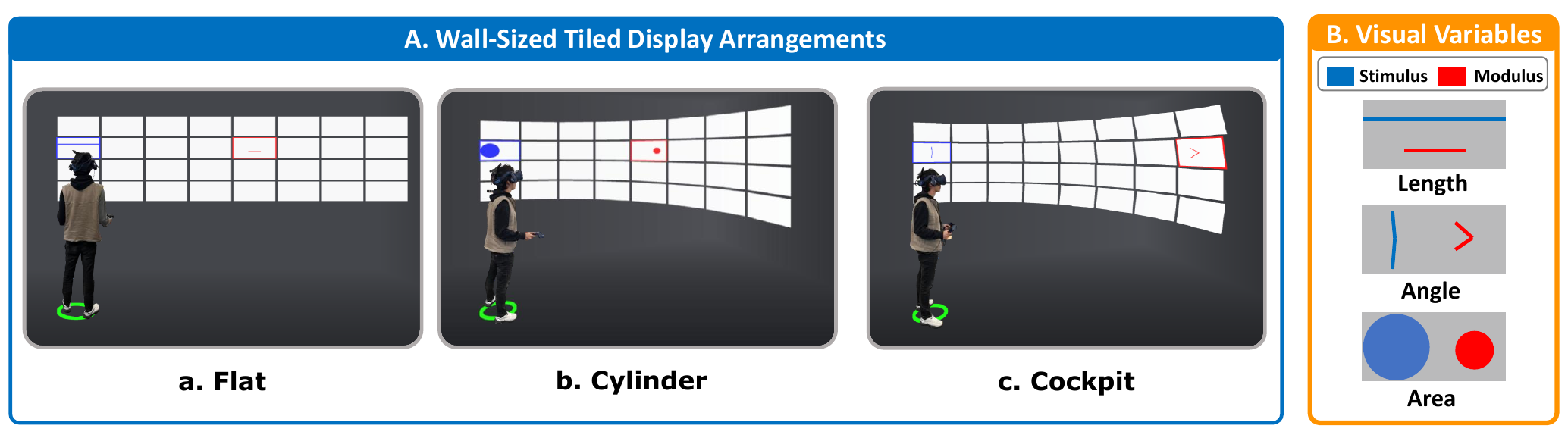}
  \caption{
  (A) The three wall-sized tiled display arrangements are investigated for magnitude reproduction tasks in an immersive environment. 
  They are flat, cylindrical, and cockpit arrangements, named \studyFactors{Flat}, \studyFactors{Cylinder}, and \studyFactors{Cockpit} in this work. 
  (B) Three visual variables including \studyFactors{Length}, \studyFactors{Angle}, and \studyFactors{Area} are used. The stimulus (blue) encodes a visual variable which participants can resize; while the modulus (red) is the target magnitude of the visual variable shown at a different location of the display. Participants are asked to resize the blue stimulus to match the red modulus as accurately as possible. 
  }
  \label{fig:teaser}
}

\graphicspath{{figs/}{figures/}{pictures/}{images/}{./}} 

\usepackage{tabu}                      
\usepackage{booktabs}                  
\usepackage{lipsum}                    
\usepackage{mwe}                       
\usepackage{helvet}

\usepackage{mathptmx}                  
\usepackage{amsmath} 

\newcommand{\revision}[1] 
{\textcolor{black}{#1}}

\newcommand{\studyFactors}[1] 
{\textit{#1}}

\newcommand{\studyMeasure}[1] 
{#1}

\newcommand{\ciValues}[1] 
{\small{#1}\normalsize{}}

\begin{document}


\firstsection{Introduction}

\maketitle
Immersive analytics is a rapidly advancing field within virtual reality (VR) and augmented reality (AR), dedicated to performing data visual analytics in immersive environments~\cite{dwyer2018immersive, liu2023datadancing}. 
Compared to traditional 2D display workspaces, immersive 3D spaces offer a larger information space surrounding individuals, allowing them to browse and organize information more effectively~\cite{lisle2021sensemaking}.
However, a key aspect of data visualization is the precise assessment and comparison of visual variables.
Visual variables~\cite{munzner2014visualization} refer to graphical attributes used to encode data, such as length, area, or hue. 
For example, in a bar chart, users assess data by comparing the lengths of bars (or the positions of the top of the bars) to analyze trends and their changes.
For effective use of large 3D information spaces, understanding how accurately users can perceive and differentiate visual variables is essential.
While data visualizations can be presented in either 2D or 3D graphic forms, this work specifically focuses on virtual wall-sized tiled displays that provide large 2D information spaces/screens in VR.

Virtual wall-sized tiled displays could offer distinct advantages over their physical counterparts~\cite{shupp2009shaping}.
Unlike physical ones, they do not require expensive equipment and maintenance and can be easily resized and repositioned as needed.
However, they still retain the benefits of physical large displays, such as providing a broader perspective on information and enhancing external memory capacity~\cite{andrews2010space,shupp2006evaluation}.  
Moreover, VR introduces unique interaction possibilities that are impossible in the physical world, enabling users to select distant objects and navigate virtual environments without the need for physical movement within their real-world space~\cite{laviola20173d, han2023evaluating}.

This paper investigates the human perceptual ability to read the magnitude of visual variables on a virtual wall-sized tiled display in different configurations. 
When visualizations are presented on large displays, some visual elements may be positioned far away or at extreme angles relative to users' viewpoint, which can affect the perception of visual variables~\cite{bezerianos2013perceptual}. 
Previous research has demonstrated that the accuracy in perceiving visual variables varies with different viewing angles on physical wall-sized tiled displays\cite{bezerianos2012perception}. However, it remains unclear whether these findings apply to virtual settings. 
Given that current VR headsets offer relatively low resolution, the perception of visual variables on virtual wall-sized tiled displays could be adversely affected. 
This work aims to explore this issue and evaluate the effectiveness of immersive analytics in utilizing virtual wall-sized tiled displays.


Specifically, our work is interested in three factors that may impact the perception of visual variables on virtual wall-sized tiled displays: 1) the configurations of the virtual displays, 2) the position of visual variables, and 3) the user interaction techniques available in VR. 
To this aim, we conducted two user studies. 
The first study evaluated how accurately users perceived visual variables, including length, angle, and area, displayed on a virtual display that replicated the setup of Bezerianos and Isenberg's study on physical wall displays \cite{bezerianos2012perception}. 
The second study examined the effects of interaction techniques on perceiving visual variables within VR, evaluating performance in terms of perception accuracy and completion time.
We studied four interaction techniques: \textit{Selection, Walking, Steering,} and \textit{Teleportation}. 
\revision{The contributions of this work are the following:  
\begin{enumerate}
\setlength\itemsep{0pt}
\item We present two user studies to investigate the human perception of reading visual variables on virtual wall-sized tiled displays in VR, considering two benefits of immersive environments including flexibility of virtual displays and unique user interactions.
\item We replicate a previous real-world study on wall-sized displays ~\cite{bezerianos2012perception} in VR to evaluate the potential of virtual wall-sized tiled displays, comparing our findings with the original results.
\item 
We present findings from two formal user studies, exploring the potential of virtual wall-sized tiled displays for immersive analytics and outlining directions for future work.
\end{enumerate}
}
\section{Related Work}


\subsection{Immersive Analytics and Virtual Display}

Earlier research~\cite{kraus2022immersive, etemadpour2013effect, arms1999benefits} suggested that immersive analytics reduce barriers between analysts and data through a greater sense of immersion, allowing analysts to analyze data-driven problems in a more intuitive and natural manner.
The authors proposed that a 'true' 3D representation can improve users' comprehension of 3D visualizations in 3D space, leading to more accurate distance estimation~\cite{etemadpour2013effect}, cluster identification~\cite{kraus2019impact}, and outlier detection~\cite{wagner2018immersive}.

Recent research has explored the use of virtual 2D displays as a substitute for physical displays, emphasizing that immersive analytics provides a larger information space compared to physical counterparts.
Kobayashi et al. \cite{kobayashi2021translating} conducted a user study involving experienced users of large displays. They found that participants dynamically generated and rearranged virtual displays according to their needs and preferred virtual display setups over physical ones. 
Lisle et al.~\cite{lisle2021sensemaking} also demonstrated the effectiveness of immersive environments in sensemaking from large datasets, showcasing the concept of Space to Think~\cite{andrews2010space} in VR.
In et al.~\cite{in2024evaluating} found that wider information space in VR enhanced user performance compared to the desktop when working with a computational notebook interface.
However, Chapuis et al.~\cite{chapuis2024comparing} pointed out that VR Head Mounted Displays (HMDs) still lack sufficient resolution and that users might still favor physical displays. 
While virtual displays offer many benefits, our understanding of how to effectively perform visual analytics in immersive environments remains limited.
For example, there is a lack of understanding of how accurately individuals can perform low-level perceptual tasks on virtual displays. This is the problem we are tackling in this paper.

\subsection{Benefits of Large Displays in Visual Analytics}\label{sec_rw_largeDisplayBenefits}
Andrews et al.~\cite{andrews2011information} characterize a large display as one that is of human scale---``closely matched to the sphere of perception and influence of the human body.'' 
Large displays are capable of accommodating a greater volume of information and make better use of the wider human field of view compared to smaller displays.
Previous research reported that large displays provide considerable cognitive and perceptual benefits for visualization tasks by providing users with an expansive information space~\cite{ andrews2013impact, czerwinski2003toward}. 
Its key benefits include the capacity to incorporate detailed and more graphical representations as well as the externalization of memory~\cite{ball2007move}. 
These features can improve user performance across a broad spectrum of visualization tasks including making spatial judgments~\cite{tan2003similar}, recognizing patterns~\cite{yost2007beyond}, and classifying objects~\cite{liu2014effects}.
As a result, large displays support users in sensemaking~\cite{andrews2010space, bradel2013large} and knowledge development~\cite{leigh2019usage}. 
Moreover, the use of large displays is well-suited for collaborative works, enabling the simultaneous involvement of multiple users in visualization analysis~\cite{isenberg2011co, jakobsen2014up, prouzeau2017tradeoffs}.

\subsection{Virtual Display Layouts for Immersive Analytics}

Virtual displays can be easily created, reconfigured, and arranged within an immersive space according to users' needs~\cite{luo2022should, satkowski2022above, reipschlager2020personal, James2023evaluating}. 
A design space for multiple display configurations for immersive analytics has four key dimensions~\cite{liu2020design}: \textit{display dimension, curvature, aspect ratio,} and \textit{orientation}. \textit{Display dimension} denotes the dimensionality of the display arrangement grid, in 2D and 3D. \textit{Curvature} describes the degree to which the display grid is curved. \textit{Aspect ratio} indicates the number of displays in each orthogonal dimension. \textit{Orientation} refers to the relative orientations of displays to users. 
Based on the design space, the most common display configurations in the literature are cylindrical, spherical, and flat surface layouts. The cylindrical and spherical layouts position displays around the user in a cylindrical or spherical manner, with prefixes such as "semi-" or "half-" and "full-" indicating the extent to which they wrap around the user. 
The flat surface layout is a conventional arrangement where multiple displays are positioned on a single plane. 
Previous research has found that placing data visualizations in cylindrical and spherical layouts is preferable to the flat layout. 
For example, Liu et al.~\cite{liu2022effects} compared these three layouts in tasks involving reading 3D bar charts.
They found no clear differences in performance across the layouts, but participants preferred the half-cylindrical layout. 
Satriadi et al.~\cite{satriadi2020maps} and Lisle et al.~\cite{lisle2021sensemaking} also suggested the use of spherical arrangements to organize multiple maps and text datasets in VR. 
As a result, our study adopts this design space for arranging virtual wall-sized tiled displays, using the three layouts with the flat surface as the baseline condition.

\subsection{Visual Variables and Perception}\label{subsec_relatedWork_perception}
How effectively individuals perceive visual variables is an important criterion in a visualization design process~\cite{wigdor2006effects}. 
Previous research in the field of data visualization has identified that the perceived magnitude of visual variables depends on factors such as the viewing distance and viewing angle between the visual variables and the user, as well as the size of the variables. 
Earlier studies conducted in real-world settings explored the impact of these factors across various display types and user positions. However, the findings across studies are not entirely consistent. 
Cleveland and McGill~\cite{cleveland1984graphical} introduced 10 visual variables, named elementary perceptual tasks, including length, angle, and area. 
Drawing on psychophysical evidence, they hypothesized that perceiving length and angle were more accurate than perceiving area while perceiving length and angle were comparable (length $\approx$ angle $<$ area)~\cite{mackinlay1985expressiveness}. Wigdor et al.~\cite{wigdor2007perception} investigated the accuracy of visual variable judgment using tabletop displays and found that length perception was the most accurate, followed by angle and area perception (length $<$ angle $<$ area).
Bezerianos and Isenberg~\cite{bezerianos2012perception} conducted perception studies using large, wall-sized tiled displays. 
Their results revealed that length perception was the most accurate, while angle perception was the least  (length $<$ area $<$ angle).
They also observed that viewing angle greatly impacted accuracy, especially when users were close to the displays and the observed object was far from them. They confirmed that observing visual variables from a sufficient distance from the displays substantially improved performance. This approach was as effective as physically moving around the large display environment. 
In this work, we investigate users' performance in perceiving visual variables on virtual wall-sized tiled displays and compare our study results with those of Bezerianos and Isenberg, seeking to understand the differences in accuracy between virtual and real-world settings.

\subsection{Perception in VR}\label{subsec_relatedWork_VRperception}

Understanding user perception and spatial awareness in VR is crucial for accurately replicating real-world experiences within virtual environments~\cite{ries2009analyzing}. 
Earlier studies showed that individuals often misjudged distances in VR, either underestimating or overestimating them compared to the real-world ones~\cite{henry1993spatial, loomis2003visual, interrante2006distance}. These studies indicate that several factors can lead to more accurate distance perception. For example, using a high-fidelity avatar and virtual environments can improve spatial perception by referencing familiar objects or one's body parts (e.g., hands)~\cite{choudhary2021revisiting, phillips2010avatar}. The position of virtual objects relative to the user's eye level can influence perception~\cite{leyrer2015eye}. 
\revision{For further details on human perception in VR, please refer to their study settings and procedures.}
In this work, we designed our virtual environment as an empty, dark room to investigate human perception abilities for the following reasons. 
\revision{First, we aimed to minimize potential confounding effects, such as distractions from other virtual objects. Kiluk et al.~\cite{kiluk2023impact} reported that participants experienced fewer distractions and great task focus in empty and dark environments.
Second, Erickson et al.~\cite{erickson2020effects} emphasized that dark environments can help reduce eye fatigue.}
We also chose to display only controllers, without avatars, and adjusted the position of the virtual wall-sized tiled display to align with the users' height.

\section{Motivation and Methods}

\revision{To effectively utilize virtual displays for immersive analytics, it is crucial to investigate their impact and limitations on the perception of visually encoded data.
 This investigation is essential, as data visualization frequently demands precise assessment and comparison of data represented in graphical forms~\cite{munzner2014visualization}. 
This study aims to explore human perceptual abilities in reading visual variables and to provide valuable insights into the effectiveness of virtual displays and 3D user interactions, highlighting the following advantages of immersive technologies:}

\begin{itemize}[leftmargin=0.15in]
\item 
\textbf{Flexibility in creating and placing virtual displays: }
The display format, position, and size can be freely customized in VR~\cite{kobayashi2021translating, lisle2021sensemaking}. 
Additionally, users can maintain a personal display in close proximity while examining a large virtual display~\cite{liu2023datadancing}.


\item \textbf{User interactions unattainable in the real world: } Immersive environments allow interactions that are impractical in the physical world~\cite{al2018virtual, argelaguet2013survey}.  
Users can explore virtual space without physical movement by employing teleportation~\cite{bowman1997travel} and steering~\cite{ruddle2009benefits}.
In addition, they can interact with distant objects by using dedicated distant manipulation techniques~\cite{bowman20043d, han2022portal}.
\end{itemize}

\subsection{Apparatus}
We used a Vive Pro Eye headset with a Vive Wireless Adapter and two controllers, providing a 110\textdegree{} field of view, 1440 \texttimes 1600 resolution per eye, and a 90 Hz refresh rate (Appendix \ref{appendix_vr_device}).
Both controllers had the same physical design, with a trigger and a circular touchpad button, but they served different functions.
The ``task-performing controller'' was always held in the dominant hand. 
The trigger button was used to initiate or finalize a task trial, while the touchpad button was used to adjust the size of visual variables during the task.
The ``interaction controller'' was held in the non-dominant hand in Study 2 (interaction details will be presented in Section~\ref{sec_study2}). The task-performing controller was used in both Study 1 and 2, while the interaction controller was used only in Study 2. To distinguish them in VR, a compass mark was affixed to the touchpad button of the interaction controller. 
We developed and ran our virtual environment for user studies using Unity 2021.3.17f1 on a Windows 11 desktop with an Intel Xeon W-2245 CPU (3.90GHz), 64GB RAM, and Nvidia GeForce RTX 3090 graphics card.

\subsection{Virtual Wall-Sized Tiled Display}

Although virtual displays can be freely positioned within immersive environments, this work focuses on investigating the effectiveness of utilizing virtual wall-sized tiled displays.
We chose the wall-sized tiled display layout because it is beneficial not only for individual use but also for collaborative purposes~\cite{belkacem2022interactive}. Hereinafter, we refer to virtual wall-sized tiled displays as \textbf{``virtual displays.''}
As illustrated in \autoref{fig:wall-setting}, our virtual display consisted of 32 individual displays arranged in a 4 by 8 grid. 
It measured approximately 5.5m $\times$ 1.8m (width $\times$ height) in total. Following chess notation, we name columns from left to right as A$\sim$H (\textbf{\studyFactors{Col\textsubscript{A}}} $\sim$ \textbf{\studyFactors{Col\textsubscript{H}}}), and the rows from bottom to top as 1$\sim$4. Each display was implemented as a 32-inch 16$\times$ 10 widescreen display, with dimensions of 65cm$\times$ 41cm (width $\times$ height), excluding the bezel dimensions. Considering a bezel size of 2cm, each display was spaced 4cm apart. 
In this work, leveraging the advantages of VR, the third row was positioned at the participants' eye level.

Our virtual displays incorporate ``bezels'' (a grid) similar to those that physical wall-sized tiled displays, herein referred to as \textbf{``physical displays''}, often have. As shown in Fig~\ref{fig:wall-setting}, the empty space between the virtual display units functions as the bezels. 
\revision{This design choice was consciously made for several reasons, even though VR allows the elimination of physical constraints such as bezels.
First, from an experimental perspective, including bezels allowed us to accurately replicate the study of Bezerianos and Isenberg ~\cite{bezerianos2012perception} on the perception of visual variables on physical wall displays and compare the results. 
Second, bezels act as visualization grid lines that define the dimensions of a visualization. Providing such grid lines (e.g., x- and y-axes) is a common practice~\cite{munzner2014visualization}. Bezerianos and Isenberg also highlighted that bezels on wall-sized displays can serve as grid lines, potentially enhancing the accuracy of interpreting basic visual variables.
Lastly, depth and distance perception in VR is not necessarily the same as in the physical world \cite{el2019survey}. Depth cues are needed to help users determine the distance, size, or even curvature of the virtual displays~\cite{liu2020design}, and the bezels serve as visual reference points to support these tasks.
}


\subsection{Magnitude reproduction Task in VR}

Our VR study (Utah State University IRB \#13757) employed a magnitude reproduction task, following study methodologies from prior psychophysics research, to investigate the relationship between physical stimuli and their perceptual interpretations~\cite{buchsbaum1971neural}. 
The task involved matching the stimulus magnitude (\studyFactors{Length}, \studyFactors{Angle}, or \studyFactors{Area} displayed in blue, see \autoref{fig:teaser}B) to the modulus magnitude (displayed in red). 
In essence, the modulus magnitude signifies the true magnitude, whereas the adjusted stimulus magnitude, defined by the participant, reflects the participant's perceived modulus magnitude.

Before starting each task trial, each participant was instructed to stand in the predetermined starting position (green circle in \autoref{fig:wall-setting}), with neither the modulus nor the stimulus initially visible. 
This starting point was 3.2m from the leftmost column of the virtual displays (identified as \studyFactors{Col\textsubscript{A}} in \autoref{fig:wall-setting}). 
The decision to start from the leftmost column of the virtual displays was based on the expectation of symmetrical visual distortion to both the left and right, centered around \studyFactors{Col\textsubscript{A}}, while also allowing for the testing of the most extreme distances. 
To assist participants in locating where the stimulus and modulus would appear on the virtual displays before the task began, reducing search time, the bezels of two virtual displays were colored blue and red, respectively (\autoref{fig:teaser}A). 

Once the participant was in the starting position and pressed the trigger button on the task-performing controller, the task began, and the stimulus and modulus appeared.
To minimize the potential impact of bezels~\cite{ball2005analysis, bi2010effects}, each stimulus and modulus were rendered entirely within individual display tiles, avoiding any overlap across bezels (tile boundaries). 
The positions of stimulus and modulus varied depending on Study 1 and 2 conditions (Section~\ref{sec_study1} and \ref{sec_study2}).
The participant could adjust the stimulus magnitude using the touchpad button on the same controller. 
Pressing down the touchpad's upper semicircle enlarged and its lower semicircle reduced the stimulus magnitude. 
The participant could accelerate stimulus magnitude changes by pressing near the top or bottom edge of the touchpad, while pressing closer to the center allowed for finer, more precise scaling.
When the participant judged that the stimulus and modulus matched in magnitude, he or she could press the trigger again to finalize the trial and proceed to the next trial.

\subsection{Measurements and Analysis}
For each task trial, we recorded four quantitative measurements:
\begin{itemize}[leftmargin=0.15in]
\setlength\itemsep{0pt}
\item \textbf{Task Completion Time:} the time required to complete a task.
    \item \textbf{Absolute Error (\studyMeasure{AbsErr}):} the absolute difference between the magnitude of the stimulus and the magnitude of the modulus, expressed as a percentage. It is computed as
\begin{equation*} AbsErr (\%) = |(S_{stimulus}-S_{modulus})/S_{modulus}| \times 100
\end{equation*}
\vspace{-0.5cm}
    \item \textbf{Estimation Error (\studyMeasure{EstErr}): } the difference between the magnitude of the stimulus and the magnitude of the modulus, expressed as a percentage.
    This measure represents the directional tendency of the interpretation error, where positive values indicate a tendency towards overestimation and negative values indicate underestimation. 
    It is defined as 
\begin{equation*}
    EstErr (\%) = (S_{stimulus}-S_{modulus})/S_{modulus} \times 100
\end{equation*}
\vspace{-0.7cm}
\item \textbf{User Movement:} the participant's head position during the task to analyze user movement and task strategies.
\end{itemize}
We also collected participants' preferences and self-evaluation of their performance through a questionnaire.

To analyze the results, we calculated bootstrapped 95\% confidence intervals (CI) for sample means and mean differences~\cite{calmettes2012making, dragicevic2016fair}. 
We calculated CIs using a bias-corrected and accelerated bootstrap (BCa) with 10,000 iterations. For study 1 (between-subjects), we used bootstrap CI calculations for two independent samples. All mean differences were adjusted for multiple comparisons using the Bonferroni correction~\cite{higgins2004introduction}. 
When interpreting the CIs of mean differences, a CI that does not overlap with 0 provides evidence of a difference, corresponding to statistically significant results in traditional p-value tests. Nonetheless, CIs allow for more nuanced interpretations~\cite{cockburn2020threats, dragicevic2016fair}: the evidence is stronger the farther the mean differences are from zero and the tighter their CIs.

\section{Study 1: Virtual Display Layout Comparisons}\label{sec_study1}

\revision{Study 1 investigates the perception accuracy of reading visual variables displayed on different virtual display configurations and compares our results to prior work on a physical display.
The specific research questions for Study 1 were:
\begin{enumerate}[label=\textbf{RQ\arabic*.}, leftmargin=0.35in]
\setlength\itemsep{0pt}
\item To what extent do virtual display arrangements influence users' perception of visual variables?
\item How does the perception accuracy of visual variables on a virtual display compare to that on a physical display?
\end{enumerate}
}

\subsection{Study Design Factors and Hypothesis} 
The study followed a mixed design featuring one between-subject factor and four within-subject factors. The between-subject factor was the type of virtual display arrangement. The within-subject factors are the visual variable type, the stimulus display location (that participants adjusted), and the location and magnitude of the modulus (target). 
To counterbalance the visual variables and stimulus locations, a Latin square design was used. The order of modulus location and magnitude was determined randomly.

\subsubsection{Virtual Display Arrangement}

Three wall display conditions were considered: \textbf{\studyFactors{Flat}}, \textbf{\studyFactors{Cylinder}}, and \textbf{\studyFactors{Cockpit}} (\autoref{fig:teaser}A), selected due to their extensive use in previous research~\cite{liu2022effects, satriadi2020maps}. 
They differed in curvature and orientation while sharing identical dimensions and aspect ratios. 
\studyFactors{Flat} arranged the virtual display tiles on a singular plane without curvature.
\studyFactors{Cylinder} arranged the virtual displays in a quarter-circle pattern centered around the participant, with the displays' y-axis oriented towards the participant. \studyFactors{Cockpit} was similar to \studyFactors{Cylinder}, but with the displays facing the participant along both the x and y-axes.

\subsubsection{Stimulus Location}

A stimulus location can be thought of as an ``answer'' location, where the stimulus was displayed, and the participant adjusted its magnitude to match their perception of the modulus from a distance. The study includes two stimulus location conditions: Frontal Display (\textbf{\studyFactors{Frontal}}) and Personal Display (\textbf{\studyFactors{Personal}}).

\begin{figure}[tb]
    \centering
    \includegraphics[width=.9\linewidth]{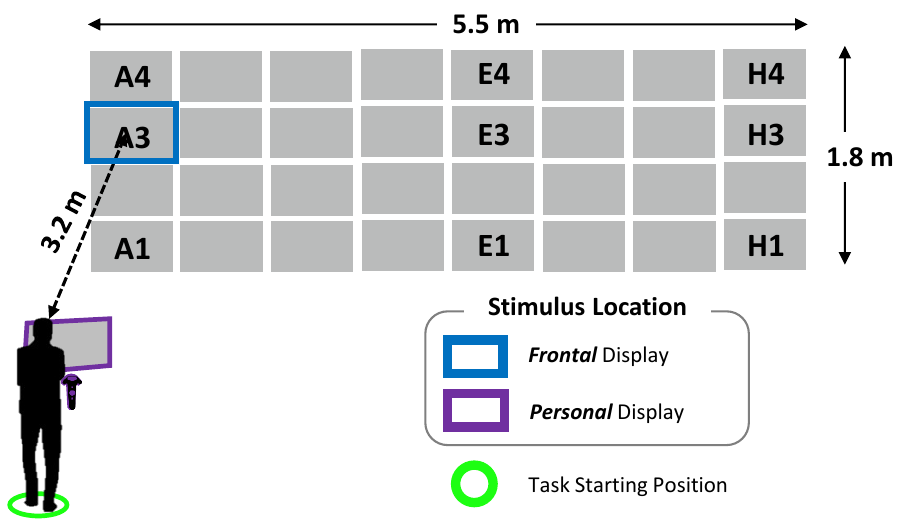}
    \caption{   
    The stimulus was shown either on a frontal display (\studyFactors{Frontal}) or a personal display (\studyFactors{Personal}). The modulus was shown in one of nine different positions on the virtual wall (A1$\sim${}H4).
    }
    \label{fig:wall-setting}
    \vspace{-0.3cm}
\end{figure}

In \studyFactors{Frontal}, the stimulus was displayed on A3 in \autoref{fig:wall-setting}, the leftmost virtual display positioned at approximately eye level.
We considered \studyFactors{Frontal} to simulate situations where the user is positioned to look straight at data on a large wall display~\cite{ball2007move, endert2011visual}.
\revision{In \studyFactors{Personal}, the stimulus display was attached to the task-performing controller, providing users with a personal display that keeps relevant information readily accessible.
It moved in sync with the user's controller movements. 
This condition was included for two key reasons. First, \studyFactors{Personal} leverages the advantage of VR (i.e., flexibility of virtual displays)~\cite{ens2014ethereal, liu2023datadancing}. 
Second, recent research has highlighted user behaviors that involve forming workspaces close to themselves during knowledge generation processes~\cite{kobayashi2021translating}.
Consequently, \studyFactors{Personal} simulates an immersive analytics scenario where users compare two visual variables—one displayed in their immediate workspace and the other on a wall-sized virtual display—before bringing the latter closer for further examination. }
In this study, the size of \studyFactors{Personal} was identical to one of the virtual display tiles to be consistent with \studyFactors{Frontal}.

\subsubsection{Visual Variables}\label{sec_visElement}

The visual variables used in our studies are \studyFactors{Length}, \studyFactors{Angle}, and \studyFactors{Area} (Figure~\ref{fig:teaser}B). 
\studyFactors{Length} was represented by a line along the x-axis, while \studyFactors{Angle} was depicted by angles on the left side. \studyFactors{Area} was measured as the area of a circle.
These variables were selected based on Cleveland and McGill's three highest-ranked elementary graphical perception tasks~\cite{bezerianos2012perception}.
The prior work found that length judgments on physical displays were relatively unaffected by viewing distance and the location of the modulus, whereas area and angle judgments were affected. We were interested to see whether these findings would hold for our virtual display settings.

When a task trial began, the initial magnitude of stimulus for \studyFactors{Length}, \studyFactors{Angle}, and \studyFactors{Area} was set to 65 cm for length (i.e., the width of one single display), 178 degrees, and 41 cm in diameter (i.e., the height of one single display). 
The choice of 178 degrees, which is slightly smaller than 180 degrees, was intended to guide the participant in understanding the direction in which the angle would change based on their controller operation.
Through controller operation, participants could adjust \studyFactors{Length} and \studyFactors{Area} diameter by a maximum of + or -0.25 cm per frame-time (0.02 seconds), while \studyFactors{Angle} was adjustable by + or -1 degree per frame-time.

\subsubsection{Modulus Location and Magnitude}

The modulus locations followed prior research \cite{bezerianos2012perception}, with a modulus appearing at one of nine specific intersections across columns A, E, and H  with rows 1, 3, and 4 (\autoref{fig:wall-setting}). 
The modulus had one of three magnitudes: 0.1, 0.4, or 0.7 times the initial stimulus magnitudes, referred to as \studyFactors{0.1S}, \studyFactors{0.4S}, and \studyFactors{0.7S}, respectively. For \studyFactors{Angle}, these multipliers were applied to 180 degrees.
These magnitudes were selected because previous research demonstrated that they led to clear differences in participants' judgments.

\begin{figure*}[tb]
    \centering
    \includegraphics[width=.97\linewidth]{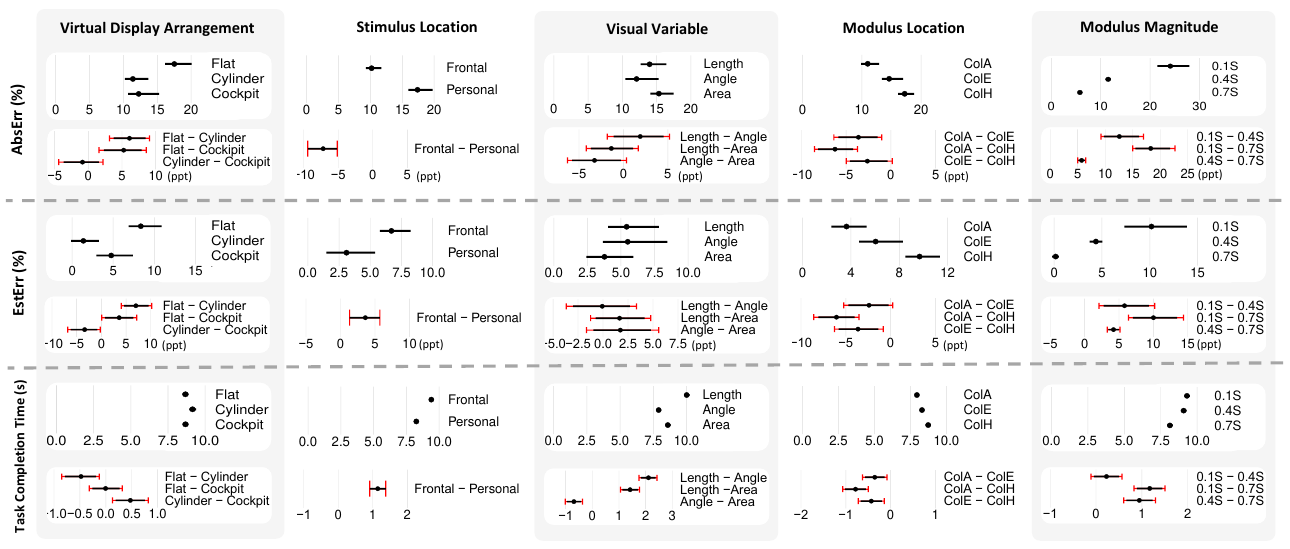}
    \caption{
    Study 1 analysis results. For each set of results by rows and columns, the measurement averages are presented first, followed by the pairwise comparison results with error bars. The error bars represent 95\% Bootstrap confidence intervals (CIs). Adjusted CIs for the pairwise comparisons with Bonferroni correction are highlighted in red.
    }
    \label{fig:s1-result}
    \vspace{-0.5cm}
\end{figure*}

\subsubsection{Hypotheses}

The hypotheses for Study 1 were as follows:
\begin{enumerate}[label=\textbf{H\arabic*.}]
\setlength\itemsep{0pt}

\item Participants' judgments are expected to yield a lower \studyMeasure{AbsErr} on  \studyFactors{Cylinder} and \studyFactors{Cockpit} than \studyFactors{Flat}. 
This is because the modulus will be shown closer to the viewer and at less acute viewing angles in \studyFactors{Cylinder} and \studyFactors{Cockpit} compared to \studyFactors{Flat}. 

\item Participants complete the task quickly in \studyFactors{Flat} compared to  \studyFactors{Cylinder} and \studyFactors{Cockpit}. This is because there is less need for head and body rotation to read a modulus in \studyFactors{$Col_{E}$} and \studyFactors{$Col_{H}$}.

\item Participants will have lower accuracy with \studyFactors{Personal} than \studyFactors{Frontal}. This is because a stimulus displayed on \studyFactors{Personal} will have a larger perceived magnitude difference from the modulus than those shown on  \studyFactors{Frontal} (i.e., on the virtual wall).

\item Participants will complete the tasks faster with \studyFactors{Personal} than with \studyFactors{Frontal} since they can reposition the personal display, allowing them to view both the stimulus and the modulus within their line of sight, resulting in minimal head movement.

\end{enumerate}

\noindent The following hypotheses are grounded in the findings reported by Bezerianos and Isenberg~\cite{bezerianos2012perception}.

\begin{enumerate}[label=\textbf{H\arabic*.}]
\setcounter{enumi}{4}
\setlength\itemsep{0pt}
\item \studyFactors{Length} has the smallest \studyMeasure{AbsErr}, followed by \studyFactors{Area} and \studyFactors{Angle}.

\item Participants overestimate the magnitude of the three visual variables (\studyMeasure{EstErr} will be positive), and this overestimation will increase as the modulus is located further from \studyFactors{$Col_{A}$}.

\item \studyMeasure{AbsErr} increases as the modulus is located further away from \studyFactors{$Col_{A}$}.

\item \studyMeasure{AbsErr} increases as the modulus magnitude decreases, as this requires longer controller operation by participants.
\end{enumerate}

\subsection{Participants}\label{sec_study1_participant}

We recruited 51 participants (28 females, 23 males; average age: 20.4 (18$\sim$52)) through our university's SONA system. 
They had (corrected) 20/20 vision and no physical disabilities that may affect the use of the VR devices. 
Two participants withdrew due to VR sickness in the \studyFactors{Cylinder} condition, leaving 49 participants available for analysis. 
In a between-subjects design,  participants were distributed as follows: 17 to \studyFactors{Flat}, 16 to \studyFactors{Cylinder}, and 16 to \studyFactors{Cockpit}.
Each participant received 2.0 SONA credits for an hour of participation. 
38 participants reported having prior VR experience.

\subsection{Procedure}

Upon arrival, a participant signed an informed consent form and completed a demographic questionnaire. Then, the participant was briefed on the research objectives, procedures, tasks, and the operation of VR controllers. Only the task-performing controller was used in this study. 
The participant's interpupillary distance (IPD) was measured using the Eye Measure app \cite{EyeMeasure}, to ensure clear vision through the VR headset.
Adjustments were made to the headset's lens distance using the IPD knob on the VIVE Pro Eye. 
Before the main task session, the participant engaged in a training session to become acquainted with the task and controller operations in VR.

During the main session, each participant performed the magnitude reproduction task under one of three virtual display arrangement conditions (i.e., \textit{Flat, Cylinder,} and \textit{Cockpit}). 
The main session consisted of a total of 162 task trials (9 modulus locations \texttimes{} 3 modulus magnitudes \texttimes{} 3 visual variables \texttimes{} 2 stimulus locations).
The main session was divided into two sub-sessions based on the stimulus locations. After completing each sub-session, the participant completed a subjective performance questionnaire. At the end of the study, the participant responded to a post-questionnaire, sharing their preferred stimulus locations and reasons for the choice.

\subsection{Results}\label{sec_study1_result}

A summary of the results is illustrated in Figure~\ref{fig:s1-result}.  
To make the analysis of the modulus location factor more manageable, this section reports results by columns (\studyFactors{$Col_{A}$}, \studyFactors{$Col_{E}$}, and \studyFactors{$Col_{H}$}).
For more details, please refer to our supplementary material (Appendix \ref{appendix_study1_results}).

\subsubsection{AbsErr}

\paragraph{Virtual Display Arrangement:} The mean \studyMeasure{AbsErr} results were: \studyFactors{Flat}=17.5\% \ciValues{[16.1, 20.0]}, \studyFactors{Cylinder}=11.4\% \ciValues{[10.2, 13.6]}, and \studyFactors{Cockpit}=12.2\% \ciValues{[10.7, 15.3]}. 
There was evidence that both \studyFactors{Cylinder} and \studyFactors{Cockpit} were more accurate than \studyFactors{Flat}. 
\studyFactors{Cylinder} was 6.22ppt (percentage point) \ciValues{[4.05, 8.75]} lower than that of \studyFactors{Flat}, while \studyFactors{Cockpit} was 5.38ppt  \ciValues{[2.56, 8.01]} lower than \studyFactors{Flat}. 
We found no clear evidence of a difference between \studyFactors{Cylinder} and \studyFactors{Cockpit} (0.86ppt \ciValues{[-1.58, 3.75]}).

\paragraph{Stimulus Location:} 
\studyFactors{Frontal} (10.2\% \ciValues{[9.30, 11.7]}) had a clearly lower \studyMeasure{AbsErr} than \studyFactors{Personal} (17.4\% \ciValues{[15.9, 19.8]}) by 7.19ppt \ciValues{[5.24, 9.62]}.

\paragraph{Visual Variable:} Pairwise comparisons did not provide evidence of a difference between visual variables (\studyFactors{Length}=13.9\% \ciValues{[12.7, 16.4]}, \studyFactors{Angle}=12.1\% \ciValues{[10.4, 15.3]}, and \studyFactors{Area}=15.3\% \ciValues{[14.1, 17.5]}.)

\paragraph{Modulus Location:} The results by the column locations were \studyFactors{$Col_{A}$}=11.0\% \ciValues{[9.89, 12.9]}, \studyFactors{$Col_{E}$}=14.6\% \ciValues{[13.4, 17.0]}, and \studyFactors{$Col_{H}$}=17.2\% \ciValues{[16.1, 18.8]}. There was evidence that \studyFactors{$Col_{A}$} had a lower \studyMeasure{AbsErr} than both \studyFactors{$Col_{E}$} (3.61ppt \ciValues{[1.13, 6.46]}) and \studyFactors{$Col_{H}$} (6.24ppt \ciValues{[3.74, 8.45]}). While we found no clear evidence of a difference between \studyFactors{$Col_{E}$} and \studyFactors{$Col_{H}$} (2.62ppt \ciValues{[-0.38, 4.89]}), our results suggest a potential trend indicating that \studyFactors{$Col_{E}$} may have a lower \studyMeasure{AbsErr} average compared to \studyFactors{$Col_{H}$}.

\paragraph{Modulus Magnitude:} The results showed that the smaller the modulus magnitude, the larger the error: 0.1S=24.1\% \ciValues{[21.5, 27.9]}, 0.4S=11.5\% \ciValues{[11.0, 12.1]}, and 0.7S=5.79\% \ciValues{[5.53, 6.08]}. 
Pairwise comparisons provided evidence that 0.1S had higher errors than 0.4S and 0.7S by 12.6ppt \ciValues{[9.98, 16.5]} and 18.3ppt \ciValues{[15.7, 22.3]}, respectively. 
Additionally, 0.4S had fewer errors than 0.7S by 5.71ppt \ciValues{[5.22, 6.26]}.

\textbf{In summary}, the results support that \studyFactors{Cylinder} and \studyFactors{Cockpit} had lower \studyMeasure{AbsErr} than \studyFactors{Flat} (\textbf{H1}), \studyFactors{Frontal} had a lower \studyMeasure{AbsErr} than \studyFactors{Personal} (\textbf{H3}), and \studyMeasure{AbsErr} increased as the modulus magnitude decreased (\textbf{H8}). We also found partial evidence to support \textbf{H7}, which stated that \studyMeasure{AbsErr} increased as the modulus was located further away from \studyFactors{$Col_{A}$}.
However, we found no clear \studyMeasure{AbsErr} difference between the three visual variables, leading to the rejection of \textbf{H5}.

\subsubsection{EstErr}

\paragraph{Virtual Display Arrangement:} The \studyMeasure{EstErr} results were \studyFactors{Flat} = 8.36\% \ciValues{[6.89, 10.9]}, \studyFactors{Cylinder} = 1.37\% \ciValues{[-0.11, 3.25]}, and \studyFactors{Cockpit} = 4.75\% \ciValues{[2.96, 7.41]}. Their positive mean values indicate a tendency to overestimate. 
Pairwise comparisons provided evidence that \studyFactors{Cylinder} had a lower \studyMeasure{EstErr} than \studyFactors{Flat} by 7.03ppt \ciValues{[4.37, 10.5]}. 
We also found that \studyFactors{Cockpit} had a slightly lower \studyMeasure{EstErr} than \studyFactors{Flat} by a difference of 3.52ppt \ciValues{[0.27, 7.06]}.
At last, there was some evidence that \studyFactors{Cylinder} had a lower \studyMeasure{EstErr} than \studyFactors{Cockpit} by 3.38ppt \ciValues{[0.21, 6.92]}.

\paragraph{Stimulus Location:} \studyFactors{Frontal} had an \studyMeasure{EstErr} of 6.69\% \ciValues{[5.77, 8.23]} and \studyFactors{Personal} had an \studyMeasure{EstErr} of 3.10\% \ciValues{[1.48, 5.39]}. The tendency to overestimate was present in both stimulus locations.
\studyFactors{Frontal} had a lower \studyMeasure{EstErr} than \studyFactors{Personal} by 3.59ppt \ciValues{[0.60, 6.13]}.

\paragraph{Visual Variable:} 
The \studyMeasure{EstErr} means were \studyFactors{Length}=5.42\% \ciValues{[4.03, 7.82]}, \studyFactors{Angle}=5.50\% \ciValues{[3.65, 8.45]}, and \studyFactors{Area} = 3.77\% \ciValues{[2.42, 5.92]}. 
We observed a tendency to overestimate, but no significant difference in the magnitude of this overestimation.

\paragraph{Modulus Location:} \studyFactors{$Col_{A}$} had an \studyMeasure{EstErr} of 3.64\% \ciValues{[2.37, 5.31]}, \studyFactors{$Col_{E}$} had 6.04\% \ciValues{[4.68, 8.29]}, and \studyFactors{$Col_{H}$} had 9.68\% \ciValues{[8.50, 11.4]}. 
We observed a tendency for \studyMeasure{EstErr} to increase as the modulus was further away from the participant.
Pairwise comparisons provided evidence of differences between \studyFactors{$Col_{A}$} and \studyFactors{$Col_{H}$} (6.04ppt \ciValues{[3.61, 8.49]}) as well as \studyFactors{$Col_{E}$} and \studyFactors{$Col_{H}$} (3.64ppt \ciValues{[0.72, 6.08]}). However, no clear evidence of a difference between \studyFactors{$Col_{A}$} and \studyFactors{$Col_{E}$} (2.40ppt \ciValues{[-0.07, 5.43]}) was found.

\paragraph{Modulus Magnitude:} The results were 0.1S=10.2\% \ciValues{[7.35\%, 13.9\%]}, 0.4S=4.36\% \ciValues{[3.70, 5.04]}, and 0.7S=0.16\% \ciValues{[-0.20, 0.50]}. Pairwise comparisons showed evidence of differences across all conditions. The 0.7S and 0.4S conditions had lower \studyMeasure{EstErr} than 0.1S by 10.0ppt \ciValues{[6.57, 14.7]} and 5.81ppt \ciValues{[2.36, 10.4]}, respectively. Additionally, 0.7S had a lower \studyMeasure{EstErr} than 0.4S by 4.20ppt \ciValues{[3.40, 5.05]}.

\textbf{In summary}, the results revealed participants tended to overestimate across the board, with overestimation increasing as the viewing distance increased, but not significantly between the first two column locations (\studyFactors{$Col_{A}$} and \studyFactors{$Col_{E}$}). 
Thus, \textbf{H6} is partially supported.

\subsubsection{Task Completion Time}

\paragraph{Virtual Display Arrangement:} The results were \studyFactors{Flat}=8.67s \ciValues{[8.48, 8.88]}, \studyFactors{Cylinder}=9.15s \ciValues{[8.94, 9.39]}, and \studyFactors{Cockpit}=8.67s \ciValues{[8.51, 8.86]}. 
\studyFactors{Flat} and \studyFactors{Cockpit} had a shorter completion time than \studyFactors{Cylinder} by 0.37s \ciValues{[0.07, 0.67]} and 0.47s \ciValues{[0.20, 0.76]}, respectively. No difference between \studyFactors{Flat} and \studyFactors{Cockpit} (0.07s \ciValues{[-0.19, 0.33]}) was found.

\paragraph{Stimulus Location:} The completion time for  \studyFactors{Frontal} and \studyFactors{Personal} were 9.40s \ciValues{[9.23, 9.59]}, and 8.25s \ciValues{[8.12, 8.41]}. There was evidence that \studyFactors{Personal} was faster than \studyFactors{Frontal} by 0.98s \ciValues{[0.88, 1.42]}.

\paragraph{Visual Variable:} The results were \studyFactors{Length}=10.0s \ciValues{[9.81s, 10.2s]}, \studyFactors{Angle}=7.90s \ciValues{[7.72, 8.09]}, and \studyFactors{Area} = 8.58s \ciValues{[8.40, 8.80]}. 
The pairwise comparisons showed differences among them. 
\studyFactors{Angle} had a faster task completion time than \studyFactors{Length} and \studyFactors{Area}, by 2.11s \ciValues{[1.84, 2.38]} and 0.69s \ciValues{[0.43, 0.94]}, respectively. \studyFactors{Area} also had a faster task completion time than \studyFactors{Length} by 1.42s \ciValues{[1.14, 1.72]}.

\paragraph{Modulus Location:} \studyFactors{$Col_{A}$} had a task completion time of 7.95s \ciValues{[7.80, 8.11]}, \studyFactors{$Col_{E}$} had 8.30s \ciValues{[8.14, 8.46]}, and \studyFactors{$Col_{H}$} had 8.73s \ciValues{[8.56, 8.91]}. Pairwise comparisons provided evidence of differences. \studyFactors{$Col_{A}$} took less time than \studyFactors{$Col_{E}$} (0.35s \ciValues{[0.13, 0.58]}) and \studyFactors{$Col_{H}$}  (0.78s \ciValues{[0.54, 1.03]}). We also saw that \studyFactors{$Col_{E}$} took less time than \studyFactors{$Col_{H}$} (0.43s \ciValues{[0.19, 0.68]}).

\paragraph{Modulus Magnitude:} The results showed that the smaller the modulus magnitude, the longer the task completion time was in average: \studyFactors{0.1S}=9.30s \ciValues{[9.11, 9.49]}, \studyFactors{0.4S}=9.07s \ciValues{[8.88, 9.28]}, and \studyFactors{0.7S}=8.13s \ciValues{[7.93, 8.34]}. \studyFactors{0.7S} had a shorter task completion time than \studyFactors{0.1S} (1.17s \ciValues{[0.90, 1.43]}) and \studyFactors{0.4S} (0.94s \ciValues{[0.68, 1.20]}). However, there was no evidence of a difference between \studyFactors{0.1S} and \studyFactors{0.4S} (0.23s \ciValues{[-0.02, 0.47]}). 

\textbf{In summary}, the results showed that \studyFactors{Personal} had a shorter completion time than \studyFactors{Frontal}, supporting \textbf{H4}. We also found partial evidence for \textbf{H2}, suggesting \studyFactors{Flat} and \studyFactors{Cockpit} had a shorter completion time than \studyFactors{Cylinder}, but no evidence of a difference between \studyFactors{Flat} and \studyFactors{Cockpit}. However, the differences across the conditions were not very large, with variations of less than 2.5s.

\subsubsection{Subjective Measurement}
After completing each task using a combination of display arrangement and stimulus location, participants assessed the ease of task completion using a 7-point rating scale (1: Very Easy--7: Very Hard). 
First, the \studyFactors{Cockpit} group reported that \studyFactors{Frontal} (3.44 \ciValues{[2.94, 4.13]}) and \studyFactors{Personal} (3.76 \ciValues{[3.18, 4.41]}) required comparable effort. Similarly, the \studyFactors{Cylinder} group did not show evidence of a difference between \studyFactors{Frontal} (3.13 \ciValues{[2.37, 4.19]}) and \studyFactors{Personal} (3.50 \ciValues{[2.75, 4.62]}).
In contrast, the \studyFactors{Flat} group reported that the task was easier to complete with \studyFactors{Frontal} (2.5 \ciValues{[1.94, 3.56]}) than with \studyFactors{Personal} (3.47 \ciValues{[2.71, 4.53]}), with a difference of 0.90 \ciValues{[0.77, 0.97]}.


\subsection{Comparison between VR and Real-World  Results}\label{section_result_vr_rw}
\begin{figure}[t]
    \centering
    \includegraphics[width=.95\linewidth]{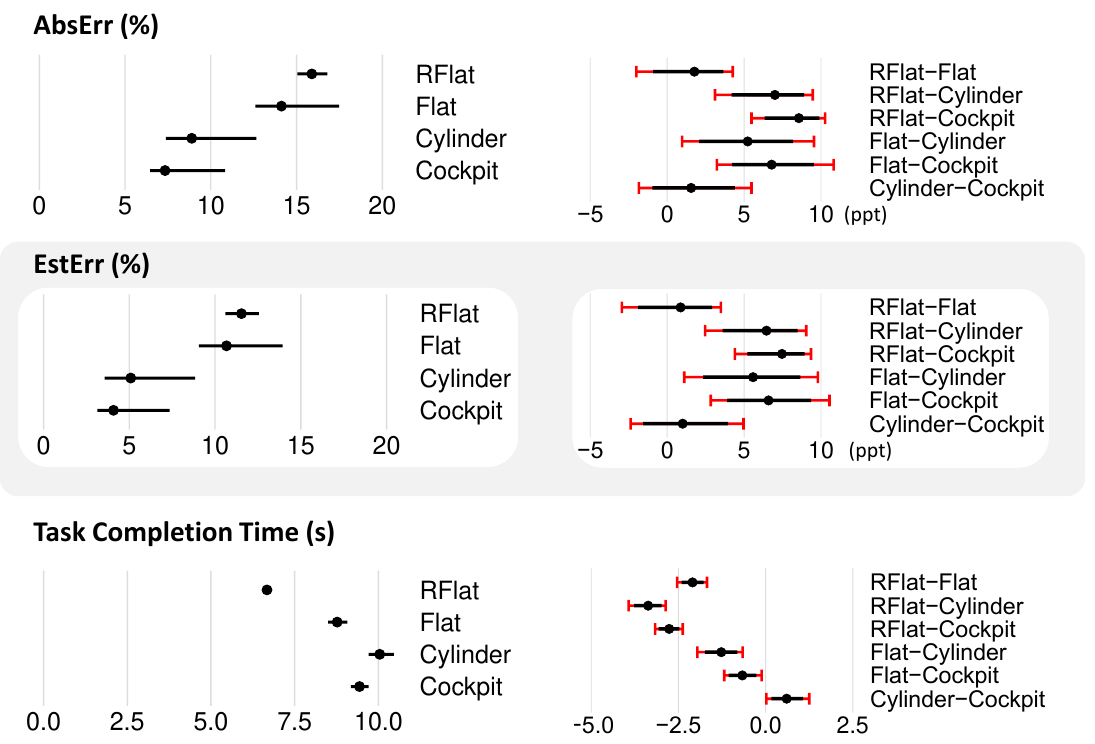}
    \caption{Comparison between our Study 1 results and the earlier work~\cite{bezerianos2012perception} investigated in the real-world setting (\studyFactors{RFlat}).
    }
    \label{fig:s1-result-rw}
    \vspace{-0.5cm}
\end{figure}

This section presents a comparative analysis of the outcomes obtained in our VR study with those investigated by Bezerianos and Isenberg~\cite{bezerianos2012perception} with a comparable physical wall. 
Please note that we can only compare a portion of their experimental results. 
Bezerianos and Isenberg tested conditions in which participants stood at 60cm and 3.2m away from the display, whereas our study focused exclusively on the 3.2m condition.
Also, the prior work was limited to perception on a flat display configuration, which is most similar to our \studyFactors{Flat} and \studyFactors{Frontal} stimulus location. We call their wall condition \studyFactors{RFlat}. 
As a result, the comparative analysis includes only outcomes of their 3.2m condition and our \studyFactors{Frontal} conditions. The results are shown in \autoref{fig:s1-result-rw}.

\subsubsection{AbsErr}
The results were \studyFactors{RFlat}=15.9\% \ciValues{[15.1, 16.8]}, \studyFactors{Flat}=14.1\% \ciValues{[12.6, 17.5]}, \studyFactors{Cylinder}=8.89\% \ciValues{[7.38, 12.7]}, and \studyFactors{Cockpit}=7.33\% \ciValues{[6.45, 10.8]}. 
Pairwise comparisons revealed some evidence of differences. 
\studyFactors{Cylinder} had a lower \studyMeasure{AbsErr} than \studyFactors{RFlat} by 6.93ppt \ciValues{[3.90, 8.68]}, and \studyFactors{Cockpit} had a lower \studyMeasure{AbsErr} than \studyFactors{RFlat} by 8.47ppt \ciValues{[6.09, 9.84]}.
There was no evidence of a difference between \studyFactors{RFlat} and \studyFactors{Flat} (1.68ppt \ciValues{[-1.11, 3.46]}).

\subsubsection{EstErr}
The results were \studyFactors{RFlat}=11.5\% \ciValues{[10.6, 12.5]}, \studyFactors{Flat}=10.7\% \ciValues{[9.06, 13.9]}, \studyFactors{Cylinder}=5.08\% \ciValues{[3.56, 8.83]}, and \studyFactors{Cockpit}=4.08\% \ciValues{[3.14, 7.34]}. 
\studyFactors{Cylinder} and \studyFactors{Cockpit} had lower \studyMeasure{EstErr} than \studyFactors{RFlat} by 6.35ppt \ciValues{[3.26, 8.21]} and 7.37ppt \ciValues{[4.92, 8.88]}, respectively. There was no clear difference between \studyFactors{RFlat} and \studyFactors{Flat} (0.87ppt [-1.89, 2.76]).

\subsubsection{Task Completion Time}
The results were \studyFactors{RFlat}=6.67s \ciValues{[6.54, 6.81]}, \studyFactors{Flat}=8.77s \ciValues{[8.50, 9.07]}, \studyFactors{Cylinder}=10.0s \ciValues{[9.71, 10.5]}, and \studyFactors{Cockpit}=9.44s \ciValues{[9.17, 9.71]}. 
\studyFactors{RFlat} was slightly faster than all the virtual ones: \studyFactors{Flat} by 1.82s \ciValues{[1.52, 2.13]}, \studyFactors{Cylinder} by 3.59s \ciValues{[3.24, 4.00]}, and \studyFactors{Cockpit} by 2.91s \ciValues{[2.62, 3.21]}.

\textbf{In summary}, the findings suggest that performance involves a trade-off between the errors and completion time. \studyFactors{Cylinder} and \studyFactors{Cockpit} showed lower \studyMeasure{AbsErr} and \studyMeasure{EstErr} but took longer to complete tasks compared to \studyFactors{RFlat}. 
We observed that participants tended to overestimate visual variables in both VR and the real-world.

\subsection{\revision{Discussion}}

In this section, we discuss our findings based on the Study 1 results and participants' comments.

\paragraph{\textbf{No evidence for differences in perception accuracy among visual variables was found in VR.}}
Our findings from Study 1 on perception accuracy among visual variables (length $\approx$ area $\approx$ angle; even when analyzing them separately by \studyFactors{Virtual Display Arrangement} as shown in Appendix \ref{appendix_study1_results} Figure~\ref{s1_flat_visVariable_error} $\sim$ Figure~\ref{s1_cockpit_visVariable_error}) are inconsistent with those from earlier real-world studies (discussed in Section~\ref{subsec_relatedWork_perception}; Cleveland and McGill~\cite{cleveland1984graphical}: length $\approx$ angle $<$ area, Wigdor et al.: ~\cite{wigdor2007perception}: length $<$angle $<$ area, and  Bezerianos and Isenberg~\cite{bezerianos2012perception}: length $<$ area $<$ angle). 
We have two hypotheses to explain this inconsistency. 
The first hypothesis is that current off-the-shelf HMDs may not yet fully replicate human visual capacity across the entire field of view~\cite{chapuis2024comparing}. \revision{The second hypothesis concerns the varying adjustment rates used to change the magnitude of visual variables, as introduced in Section~\ref{sec_visElement}.}
Further research is required to investigate these hypotheses \revision{and to explore detailed differences in perception of various visual variables beyond the three examined, in both quantitative and qualitative ways. }

\paragraph{\textbf{Virtual environments and displays could offer a practical workspace for performing visualization tasks.}} 
Comparative results in Section~\ref{section_result_vr_rw} showed that the two curved virtual displays (\studyFactors{Cylinder} and \studyFactors{Cockpit}) yielded lower \studyMeasure{AbsErr} and \studyMeasure{EstErr} than the physical displays (\studyFactors{RFlat}) for the visual variables tested, despite having the longer task completion time. 
\revision{While this finding is limited by the absence of a direct comparison between curved conditions in real and virtual environments. It indicates that virtual displays, with their flexibility in positioning, could be effectively used for visual analytics tasks.}
Nonetheless, the understanding is limited, and necessitates further exploration of diverse scenarios, particularly those involving smaller targets and more complex visualizations that may exceed the resolution capabilities of current VR headsets.
Moreover, it is crucial to consider adverse effects like VR sickness, as two participants in the \studyFactors{Cylinder} group discontinued their participation due to nausea.

\paragraph{\textbf{Supporting users in reading two visual variables at comparable depths is important.}} 
Study 1 revealed that participants were more accurate when both the stimulus and modulus were positioned on the displays (\studyFactors{Frontal}) compared to when the stimulus was on the \studyFactors{Personal} display in front of them. 
Indeed, the majority of participants (11 in the \studyFactors{Flat} group, 9 in the \studyFactors{Cylinder} group, and 12 in the \studyFactors{Cockpit} group) reported that they preferred \studyFactors{Frontal}. 
They commented that comparing the sizes of two distant objects was significantly easier than comparing sizes involving close and distant objects. 
Although we attempted to reduce this disparity by ensuring that the \studyFactors{Personal} display was the same size as the tiles on the virtual display, this adjustment proved insufficient.
These findings highlight that the accuracy and efficiency of perceiving sizes decrease as the depth disparity between the stimulus and modulus increases, indicating that a common frame of reference (e.g., depth in our study) aids in size comparison~\cite{mohler2010effect, ries2008effect}.

\paragraph{\textbf{\revision{Positioning visual variables in close proximity is an effective strategy for facilitating comparisons.}}} 
\revision{The results showed that participants' AbsErr and task completion time decreased as the stimulus and modulus were placed closer together. This finding aligns with the visualization strategy, known as juxtaposition~\cite{munzner2014visualization}, which facilitates side-by-side comparisons and proves more effective than relying on memory to recall previously viewed information.}

\section{Study 2: Effectiveness of VR Interaction}\label{sec_study2}

The objective of Study 2 was to examine the effectiveness of 3D user interaction and gain insights into user behavior during interaction tasks. \revision{It is important to note that 3D user interaction involves both navigation within the virtual environment and interaction with virtual objects.
The specific research question for Study 2 was:
\begin{enumerate}[label=\textbf{RQ\arabic*.}, leftmargin=0.35in]
\setlength\itemsep{0pt} \setcounter{enumi}{2}
\item How does the introduction of different VR interaction techniques affect the perception of visual variables? 
\end{enumerate}
}

\subsection{Study Design Factors and Hypotheses}

\begin{figure}[t]
    \centering
    \includegraphics[width=\linewidth]{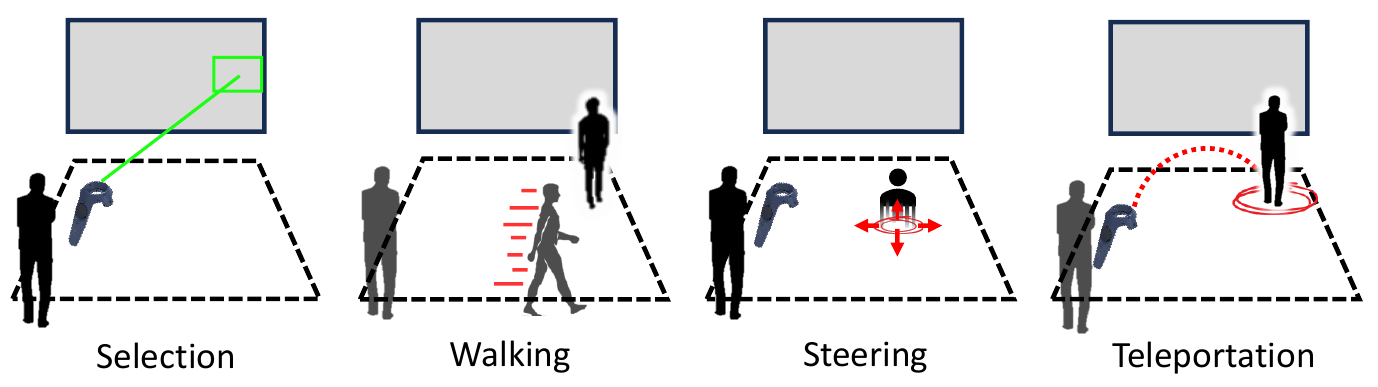}
    \caption{
    \studyFactors{Selection} is exclusive for \studyFactors{Personal}, whereas the others apply to both \studyFactors{Frontal} and \studyFactors{Personal}. \studyFactors{Selection} allows users to select a display tile and make its copy on the interaction controller. 
    \studyFactors{Steering} and \studyFactors{Teleportation} enable virtual navigation from a fixed position, while \studyFactors{Walking} requires physical movement in the real-world.
     }
    \label{fig:interaction-techniques}
    \vspace{-.5cm}
\end{figure}

Study 2 employed a within-subject design with the simplified experimental factors from Study 1 plus user interaction techniques (Fig.~\ref{fig:interaction-techniques}). 
\revision{The factors of virtual display layout, visual variables, and modulus locations were simplified to ensure that Study 2 could be completed within a manageable timeframe. 
We chose to use only the \studyFactors{Flat} layout, as it is commonly used in collaboration scenarios. 
Next, we decided to focus on a single visual variable rather than including all three, as Study 1 revealed no significant differences in AbsErr among them.
The \studyFactors{Area} variable was selected because it had the poorest average AbsErr, providing an opportunity to assess how different interaction techniques could improve accuracy.}
For modulus locations, we used three locations at the intersections of columns A, E, and H with row 3, aligning with participants' eye height~\cite{cleveland1985elements, wigdor2007perception}. 
Beyond these adjustments, we used the same stimulus locations (\studyFactors{Frontal} and \studyFactors{Personal}) and the modulus magnitudes (\studyFactors{0.1S, 0.4S,} and \studyFactors{0.7S}). 
The study took about 90 minutes.

\subsubsection{User Interaction Techniques}\label{sec_interactionTechniques}

\paragraph{No Interaction:} Participants performed tasks while standing in the same starting position as in Study 1. This condition was used exclusively for \studyFactors{Frontal} and served as a baseline for comparing other interaction techniques. \revision{For the \studyFactors{Personal} condition, please note that we decided to skip \studyFactors{No Interaction} due to its lower performance compared to \studyFactors{Frontal} in Study 1 but employ the following interaction.}

\paragraph{Selection:} 
This interaction was exclusive to the \studyFactors{Personal} condition \revision{to leverage both benefits of VR (i.e., flexibility of virtual displays and enhanced user interaction)}. 
Using ``Selection,'' participants could create a copy of the modulus display above the interaction controller held in their non-dominant hand, while the \studyFactors{Personal} display was shown on the task-performing controller.
When participants pointed the interaction controller at a display tile, a green line emerged from the controller, and the tile was highlighted by turning its bezel green. 
Pressing the trigger button copied the tile to the interaction controller.

\paragraph{Walking:} In this condition, participants physically walked around to reposition themselves in front of the virtual displays. 
This navigation method is considered the most natural, as it mirrors the biological symmetry of real-world walking.
It does not require additional controller operation, but requires a physical space in the real-world that has to match the size of the virtual environment. 

\paragraph{Steering:}
Steering allows participants to navigate the virtual environment by using the pad button on the interaction controller without physically moving in the real-world~\cite{bowman1997travel}. 
For instance, pressing the top or right part of the pad moves participants forward or to the right, relative to their current position. 
To rotate their viewpoint, participants needed to physically rotate themselves. 
In this work, its maximum speed is set to 1.50 m/s, which is comparable to the average human walking speed (1.42 m/s)~\cite{mohler2007visual, levine1999pace}.

\paragraph{Teleportation:} This technique allows participants to instantly relocate to the specified location in VR~\cite{bowman1997travel}. When participants clicked a trigger button on the interaction controller, an arc shape ray emerged from the controller to the floor. Releasing the button transported them to the targeted location with a brief fade-in and fade-out effect.
To prevent participants from losing track of the virtual display locations, their orientation was adjusted to face the displays after teleportation.

\subsubsection{Hypotheses}

Our hypotheses for Study 2 are as follows: 
\begin{enumerate}[label=\textbf{H\arabic*.}]
\setcounter{enumi}{8}
\setlength\itemsep{0pt}

\item The introduction of interactions is expected to result in lower \studyMeasure{AbsErr} compared to \studyFactors{No Interaction}, as they  allow participants to relocate themselves in locations where the difference in perceived size between the stimulus and modulus is minimized.

\item \studyFactors{No Interaction} and \studyFactors{Selection} are expected to have the shortest task completion time, as they require no or little interaction for \studyFactors{Frontal} and \studyFactors{Personal}, respectively. These are anticipated to be followed by \studyFactors{Teleportation}, \studyFactors{Steering}, and \studyFactors{Walking} based on the time required for navigation within the virtual environment.

\end{enumerate}

\subsection{Participants and Procedures}

We recruited 25 participants (8 females, 17 males; average age: 20.6 ranging from 18 to 22). 
All participants had (corrected) 20/20 vision and no disabilities affecting the use of VR devices. None had participated in Study 1. Participants received compensation of 2.5 SONA credits.
20 students reported having prior VR experience, with an average self-rated VR familiarity score of 3.9 out of 7.

While the overall procedure was almost identical to that in Study 1, Study 2 additionally asked participants to answer the VR sickness questionnaire (VRSQ)~\cite{kim2018virtual} after completing each task to measure the degree of VR sickness caused by the interaction techniques.
Each participant performed 72 task trials (4 interaction techniques \texttimes{} 3 modulus locations \texttimes{} 3 modulus sizes \texttimes{}  2 stimulus locations).

\subsection{Results}

This section focuses on investigating the effect of stimulus locations and interaction techniques on \studyMeasure{\textbf{AbsErr}} and \studyMeasure{\textbf{Task Completion Time}} to validate our hypotheses.
We present performance results for the interaction techniques and compare them to \studyFactors{No Interaction}.
For more detailed pairwise comparison results, please refer to our supplementary material (Appendix \ref{appendix_study2_results}).

\subsubsection{AbsErr}

\paragraph{Interactions and \studyFactors{Frontal}:} 
The results were \studyFactors{No Interaction}=12.8\% \ciValues{[10.8, 15.8]}, \studyFactors{Walking}=6.41\% \ciValues{[5.60, 7.43]}, \studyFactors{Steering}=5.94\% \ciValues{[5.16, 6.87]}, and \studyFactors{Teleportation}=8.36\% \ciValues{[7.03, 10.8]}. \studyFactors{No Interaction} had a higher \studyMeasure{AbsErr} than \studyFactors{Walking} (6.11ppt \ciValues{[3.27, 11.8]}) and \studyFactors{Steering} (6.55ppt \ciValues{[3.65, 11.7]}).
\studyFactors{Teleportation} had lower \studyMeasure{AbsErr} than \studyFactors{No Interaction}, but had no strong difference (4.14ppt \ciValues{[-0.51, 8.92]}). 

\begin{figure}[t]
    \centering
    \includegraphics[width=.93\linewidth]{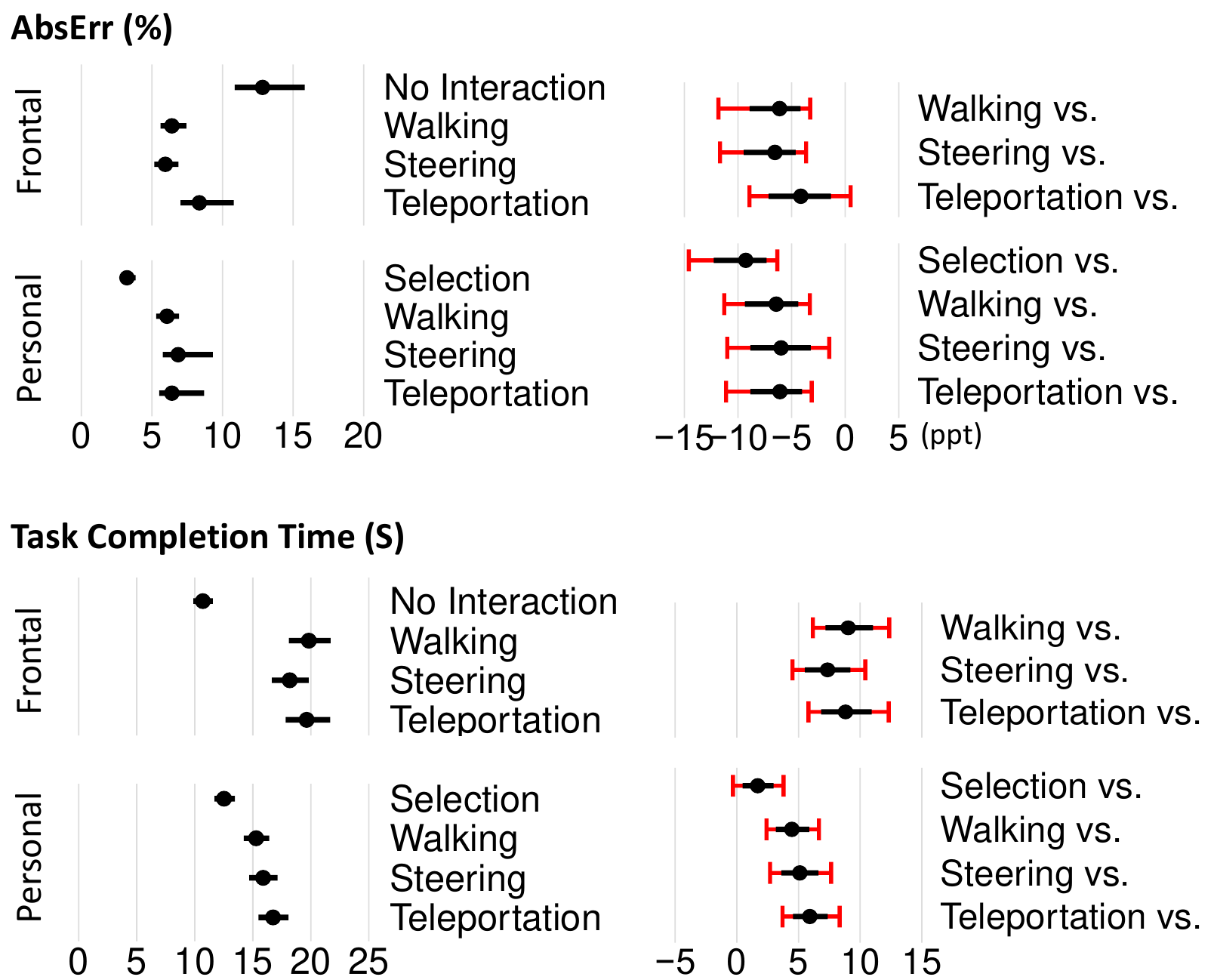}
    \caption{
    Study 2 analysis results. It shows the measurement averages and pairwise comparison results with 95\% CIs. \textit{`vs.'} indicates comparison results to the \studyFactors{No Interaction} with \studyFactors{Frontal} condition. 
     }
    \label{fig:s2-result}
    \vspace{-.5cm}
\end{figure}

\paragraph{Interactions and \studyFactors{Personal}:} 
The results were \studyFactors{Selection}=3.23\% \ciValues{[2.78, 3.84]}, Walking=6.06\% \ciValues{[5.30, 6.92]}, Steering=6.86\% \ciValues{[5.76, 9.31]}, and Teleportation=6.41\% \ciValues{[5.52, 8.69]}. 
Their pairwise comparisons to the baseline condition (\studyFactors{No Interaction} and \studyFactors{Frontal}) revealed improvement as follows: \studyFactors{Selection} by 9.27 \ciValues{[6.33, 14.6]}, \studyFactors{Walking} by 6.43 \ciValues{[3.30, 11.3]}, 
\studyFactors{Steering} by 5.98 \ciValues{[1.48, 10.98]}, and \studyFactors{Teleportation} by 6.08 \ciValues{[3.12, 11.1]}.

\textbf{In summary,} interaction lowered \studyMeasure{AbsErr}, except in the condition combining \studyFactors{Teleportation} and \studyFactors{Frontal}.
It partially supports \textbf{H9}.

\subsubsection{Task Completion Time}

\paragraph{Interactions and \studyFactors{Frontal}:} 
\studyFactors{No Interaction} had the shortest time (10.7s \ciValues{[9.87, 11.6]}), followed by \studyFactors{Steering} (18.2s \ciValues{[16.7, 19.8]}), \studyFactors{Teleportation} (19.6s \ciValues{[17.8, 21.7]}), and \studyFactors{Walking} (19.8s \ciValues{[18.1, 21.7]}). 
\studyFactors{No Interaction} was faster than \studyFactors{Steering} by 7.36s \ciValues{[4.99, 9.78]}, \studyFactors{Teleportation} by 8.81s \ciValues{[6.32, 11.6]}, and \studyFactors{Walking} by 9.03s \ciValues{[6.71, 11.7]}.

\paragraph{Interactions and \studyFactors{Personal}:} 
The average results were as follows:
\studyFactors{Selection}=12.5s \ciValues{[11.7, 13.5]}, Walking=15.3s \ciValues{[14.2, 16.4]}, Steering=15.9s \ciValues{[14.7, 17.2]}, and Teleportation=16.7s \ciValues{[15.5, 18.1]}. 
Pairwise comparisons to the baseline showed that the introduction of interactions to \studyFactors{Personal} resulted in longer completion time as follows: \studyFactors{Selection} by 1.69s [-0.31, 3.79], \studyFactors{Walking} by (4.46s \ciValues{[2.40, 6.65]}),  \studyFactors{Steering} (5.11s \ciValues{[2.69, 7.65]}), and \studyFactors{Teleportation} (5.92s \ciValues{[3.70, 8.33]}). 
However, the difference between \studyFactors{Selection} and \studyFactors{No Interaction} was not clear.

\textbf{In summary, } \studyFactors{Steering}, \studyFactors{Teleportation}, and \studyFactors{Walking} led to longer task completion times for both \studyFactors{Frontal} and \studyFactors{Personal} conditions. This finding supports \textbf{H10}.

\subsubsection{Subjective Measurement}

We first report the ease of task completion results, followed by the VRSQ results. 
For \studyFactors{Frontal}, the ease of task completion results were \studyFactors{No Interaction}=2.63 \ciValues{[2.05, 3.32]}, \studyFactors{Walking}=3.32 \ciValues{[2.63, 4.10]}, \studyFactors{Steering}=3.32 \ciValues{[2.84, 4.00]}, and \studyFactors{Teleportation}=3.42 \ciValues{[3.11, 4]}. 
Next, the results for \studyFactors{Personal} were \studyFactors{Selection}=1.79 \ciValues{[1.42, 2.47]}, \studyFactors{Walking}=3.32 \ciValues{[2.58, 4.16]}, \studyFactors{Steering}=3.32 \ciValues{[2.84, 3.95]}, and \studyFactors{Teleportation}=3.26 \ciValues{[2.68, 4.05]}. 
Next, for the VRSQ score results, please note that higher scores are less desirable. 
In \studyFactors{Frontal}, the interaction techniques had the following scores: \studyFactors{No Interaction}=14.5 \ciValues{[10.8, 20.0]}, Walking=15.3 \ciValues{[9.96, 25.1]}, Steering=20.8 \ciValues{[13.3, 31.9]}, and Teleportation=14.8 \ciValues{[10.2, 23.3]}. In \studyFactors{Personal}, the scores for the interaction techniques were \studyFactors{Selection}=10.3 \ciValues{[6.36, 17.8]}, Walking=15.0 \ciValues{[9.30, 24.5]}, Steering=18.6 \ciValues{[11.9, 27.5]}, and Teleportation=15.2 \ciValues{[8.90, 25.9]}. 
\textbf{In summary,} only \studyFactors{Selection} outperformed \studyFactors{No Interaction} in both measurements, while the others performed worse than \studyFactors{No Interaction}.

\subsection{\revision{Discussion}}

\begin{figure}[t]
    \centering
    \includegraphics[width=.91\linewidth]{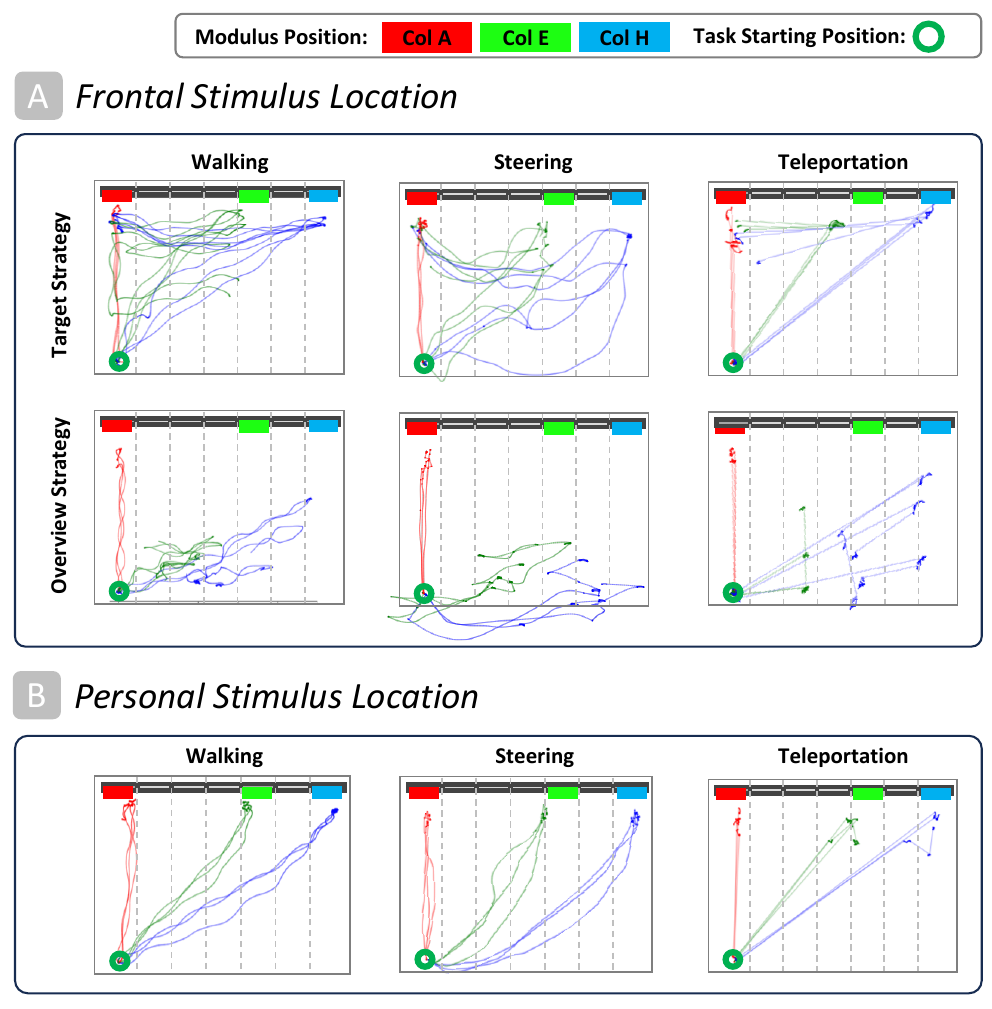}
    \caption{
     Movement strategies.
    (A) With \studyFactors{Frontal}, we observed two representative strategies: the target strategy and the overview strategy. 
    (B) With \studyFactors{Personal}, participants employed a strategy of initially moving to the modulus positions and evaluating stimulus and modulus sizes.
    }
    \label{fig:interaction-movement-result}
    \vspace{-.5cm}
\end{figure}

This section reports Study 2 findings, including movement logs and participants' comments. Figure~\ref{fig:interaction-movement-result} illustrates participants' movements when using the \studyFactors{Walking}, \studyFactors{Steering}, and \studyFactors{Teleportation} techniques.

\paragraph{\textbf{The \studyFactors{Personal} display can be enhanced through the use of 3D user interactions.}}
While Study 1 showed that \studyFactors{Personal} was less effective than \studyFactors{Frontal}, Study 2 demonstrated that 3D user interactions could improve its performance.
Moreover, 22 out of 25 participants in Study 2 preferred \studyFactors{Personal} over \studyFactors{Frontal}. 
They stated that using \studyFactors{Personal} with 3D user interactions allowed them to bring the stimulus display closer to the modulus for size comparisons, or vice versa (Figure~\ref{fig:interaction-movement-result}B), without the need to oscillate between the stimulus and modulus locations (Figure~\ref{fig:interaction-movement-result}A). 
Furthermore, participants showed an interesting strategy of utilizing Z-fighting by overlapping the stimulus and modulus to compare their sizes. Z-fighting is a phenomenon in 3D rendering where objects occupying the same spatial location cause flickering or visual instability due to rendering priority conflicts.

Conversely, two participants who preferred \studyFactors{Frontal} pointed out two issues with \studyFactors{Personal}. Notably, in this study, \studyFactors{Personal} was designed to be the same size as a single tile in virtual displays and was always attached to the task-performing controller. One participant commented that shaky hands hampered the performance of \studyFactors{Personal}, whereas \studyFactors{Frontal} remained stationary. 
This issue could be easily resolved by allowing users to relocate the stimulus display only when the controller interacts with it.
Another participant stated that \studyFactors{Personal} felt too large and often too close. This point will be discussed further in Section~\ref{sec_futureWork}.

\paragraph{\textbf{The \studyFactors{Personal} display with \studyFactors{Selection} can greatly assist users in perceiving visual variables. }} 
This finding indicates a potential advantage across different target display positions, not limited to our study setup.
However, there are some potential weaknesses. 
Because the technique involves bringing or copying a distant display to the user's side, there is a risk of losing the spatial context of virtual displays compared to \studyFactors{Frontal}. 
Furthermore, it could increase the risk of communication errors in collaborative environments. To mitigate the concerns, one possible solution is to highlight the distant display on the interaction controller after it has been copied using \studyFactors{Selection}.

\paragraph{\textbf{\studyFactors{Steering} and \studyFactors{Teleportation} demonstrate potential as alternative interaction techniques to \studyFactors{Walking}.}} Study 2 suggested that \studyFactors{Steering} and \studyFactors{Teleportation} could enhance user performance compared to the baseline (\studyFactors{No Interaction}), partially supporting \textbf{H9}.
Moreover, we observed that participants employed two similar navigation strategies, despite using different interactions, namely the overview strategy and the target strategy (Figure~\ref{fig:interaction-movement-result}A). 
The overview strategy entails initially navigating toward the center area between the modulus and stimulus before assessing their sizes. The target strategy involves oscillating between stimulus and modulus positions to assess their sizes. 
During the target strategy, nine participants interestingly attempted to use the controllers as a reference tool for comparing sizes.

The participants' comments offer valuable insights into the use of \studyFactors{Steering} and \studyFactors{Teleportation}. 
Post-questionnaire results showed \studyFactors{Walking} as the most preferred technique (12 participants), followed by \studyFactors{Steering} (5) and \studyFactors{Teleportation} (5).
Participants who preferred \studyFactors{Walking} frequently cited its naturalness as the most significant advantage, consistent with Ball et al.~\cite{ball2007move}'s findings.
In contrast, participants who preferred \studyFactors{Steering} and \studyFactors{Teleportation} valued the easy and quick movement through simple controller operations, pointing out that \studyFactors{Walking} demanded significant physical effort to move between the stimulus and modulus.

The participants' feedback also highlighted some limitations of the techniques. For \studyFactors{Steering}, three participants reported experiencing motion-related sickness. 
A potential solution to improve user experience could involve personalizing movement speed~\cite{bowman1997travel}.
For \studyFactors{Teleportation}, some participants noted that they traveled further or shorter distances than intended. 
This issue may partly stem from the virtual workspace being an empty, dark room with no additional landmarks or references beyond the virtual display.
To improve the user experience for \studyFactors{Teleportation}, adding more spatial information in the virtual environment could help users effectively execute their target or overview strategies.
Potential solutions include displaying a grid or shadows on the floor to indicate the location of the virtual displays~\cite{thompson2011visual} and providing anchors that serve as predetermined \studyFactors{Teleportation} destinations.


\section{Limitation and Future Work}\label{sec_futureWork}

Our study used a set of modulus sizes that were consistently clear and visible at all distances in VR. However, in real-world analytics, some visual elements might be smaller, potentially exceeding the rendering capability of VR headsets. The limitation is expected to be addressed in the future as HMD resolution improves at an affordable price, eventually matching human vision.

The use of virtual displays and 3D user interactions in collaborative tasks has yet to be explored.
For example, while the \studyFactors{Cylinder} and \studyFactors{Cockpit} arrangements proved effective in single-user contexts, issues may arise when multiple users occupy the same central positions, as their embodiment avatars may disrupt task execution.

We will also explore the additional benefits of virtual displays for immersive analytics.
Allowing the size of \studyFactors{Personal} stimulus display to be adjustable would help users organize multiple personal displays around them.
However, it may introduce new challenges in effective visualization design if designers cannot anticipate at what sizes their visualizations will be seen. 
For example, if visualizations are small enough, color luminance becomes a more distinct feature than hue~\cite{ware2010visual}.  
Adjusting visualization features for size could help, but it compromises interface consistency and may hinder task performance. 
We will investigate how variations in display sizes affect the user's perception ability. 
Next, the acquisition of bezel-less large displays is hindered by high costs and technical constraints in the real-world, but not in VR.
While bezels can be important reference points for reading visual variables and locating areas within large visualizations, they are generally advised against because they cause visual discontinuities~\cite{wallace2014effect}.
We aim to investigate the effects of bezel presence and size on immersive analytics.
\revision{
At last, future work should investigate the effectiveness of virtual displays in Mixed Reality (MR) environments, with a focus on visual variable perception accuracy and user interactions.
}
\section{Conclusion}

This paper examined user accuracy in reading visual variables on a virtual wall-sized tiled display across different configurations, evaluating the potential of VR environments for visualization tasks.
Study 1 showed that participants made smaller errors on virtual curved walls (\studyFactors{Cylinder} and \studyFactors{Cockpit}) compared to \studyFactors{Flat}. 
The results were compared to the previous real-world study, confirming that virtual curved walls showed better accuracy than the physical display (\studyFactors{RFlat}), though with longer task completion times.
Study 2 demonstrated that user interaction techniques can improve perception accuracy. 
The findings revealed that \studyFactors{Personal} and \studyFactors{Selection}, which are difficult to achieve in the real-world, can effectively support users' visualization tasks. 
We also discovered that \studyFactors{Steering} and \studyFactors{Teleportation} could be used as viable alternatives to \studyFactors{Walking}. 
Based on our findings, wall-sized virtual displays in VR have the potential to serve as a viable workspace for visual analytics as an alternative to real-world environments.

\newpage


\bibliographystyle{abbrv-doi-hyperref-narrow}

\bibliography{reference}

\begin{thebibliography}{10}
\renewcommand*{\sfdefault}{PTSansNarrow-TLF}

\bibitem{al2018virtual}
M.~Al~Zayer, P.~MacNeilage, and E.~Folmer.
\newblock Virtual locomotion: a survey.
\newblock {\em IEEE Transactions on Visualization and Computer Graphics}, 26(6):2315--2334, 2018. \href{https://doi.org/10.1109/TVCG.2018.2887379}
{doi: \textsf{%
10\hspace{.1pt}\discretionary{.}{%
}{.}\hspace{.4pt}1109\discretionary{/}{%
}{/}TVCG\hspace{.1pt}\discretionary{.}{%
}{.}\hspace{.4pt}2018\hspace{.1pt}\discretionary{.}{%
}{.}\hspace{.4pt}2887379}}


\bibitem{andrews2010space}
C.~Andrews, A.~Endert, and C.~North.
\newblock Space to think: large high-resolution displays for sensemaking.
\newblock In {\em Proceedings of the Conference on Human Factors in Computing Systems (CHI)}, pp. 55--64. ACM, 2010. \href{https://doi.org/10.1145/1753326.1753336}
{doi: \textsf{%
10\hspace{.1pt}\discretionary{.}{%
}{.}\hspace{.4pt}1145\discretionary{/}{%
}{/}1753326\hspace{.1pt}\discretionary{.}{%
}{.}\hspace{.4pt}1753336}}


\bibitem{andrews2011information}
C.~Andrews, A.~Endert, B.~Yost, and C.~North.
\newblock Information visualization on large, high-resolution displays: Issues, challenges, and opportunities.
\newblock {\em Information Visualization}, 10(4):341--355, 2011. \href{https://doi.org/10.1177/1473871611415997}
{doi: \textsf{%
10\hspace{.1pt}\discretionary{.}{%
}{.}\hspace{.4pt}1177\discretionary{/}{%
}{/}1473871611415997}}


\bibitem{andrews2013impact}
C.~Andrews and C.~North.
\newblock The impact of physical navigation on spatial organization for sensemaking.
\newblock {\em IEEE Transactions on Visualization and Computer Graphics}, 19(12):2207--2216, 2013. \href{https://doi.org/10.1109/tvcg.2013.205}
{doi: \textsf{%
10\hspace{.1pt}\discretionary{.}{%
}{.}\hspace{.4pt}1109\discretionary{/}{%
}{/}tvcg\hspace{.1pt}\discretionary{.}{%
}{.}\hspace{.4pt}2013\hspace{.1pt}\discretionary{.}{%
}{.}\hspace{.4pt}205}}


\bibitem{argelaguet2013survey}
F.~Argelaguet and C.~Andujar.
\newblock A survey of 3d object selection techniques for virtual environments.
\newblock {\em Computers \& Graphics}, 37(3):121--136, 2013. \href{https://doi.org/10.1016/j.cag.2012.12.003}
{doi: \textsf{%
10\hspace{.1pt}\discretionary{.}{%
}{.}\hspace{.4pt}1016\discretionary{/}{%
}{/}j\hspace{.1pt}\discretionary{.}{%
}{.}\hspace{.4pt}cag\hspace{.1pt}\discretionary{.}{%
}{.}\hspace{.4pt}2012\hspace{.1pt}\discretionary{.}{%
}{.}\hspace{.4pt}12\hspace{.1pt}\discretionary{.}{%
}{.}\hspace{.4pt}003}}


\bibitem{arms1999benefits}
L.~Arns, D.~Cook, and C.~Cruz-Neira.
\newblock The benefits of statistical visualization in an immersive environment.
\newblock In {\em Proceedings IEEE Virtual Reality (Cat. No. 99CB36316)}, pp. 88--95. IEEE, 1999.

\bibitem{ball2005analysis}
R.~Ball and C.~North.
\newblock Analysis of user behavior on high-resolution tiled displays.
\newblock In {\em IFIP Conference on Human-Computer Interaction}, pp. 350--363. Springer, 2005. \href{https://doi.org/10.1007/11555261_30}
{doi: \textsf{%
10\hspace{.1pt}\discretionary{.}{%
}{.}\hspace{.4pt}1007\discretionary{/}{%
}{/}11555261\_30}}


\bibitem{ball2007move}
R.~Ball, C.~North, and D.~A. Bowman.
\newblock Move to improve: promoting physical navigation to increase user performance with large displays.
\newblock In {\em Proceedings of the SIGCHI conference on Human factors in computing systems}, pp. 191--200, 2007. \href{https://doi.org/10.1145/1240624.1240656}
{doi: \textsf{%
10\hspace{.1pt}\discretionary{.}{%
}{.}\hspace{.4pt}1145\discretionary{/}{%
}{/}1240624\hspace{.1pt}\discretionary{.}{%
}{.}\hspace{.4pt}1240656}}


\bibitem{belkacem2022interactive}
I.~Belkacem, C.~Tominski, N.~M{\'e}doc, S.~Knudsen, R.~Dachselt, and M.~Ghoniem.
\newblock Interactive visualization on large high-resolution displays: A survey.
\newblock In {\em Computer Graphics Forum}, p. e15001. Wiley Online Library, 2022.

\bibitem{bezerianos2012perception}
A.~Bezerianos and P.~Isenberg.
\newblock Perception of visual variables on tiled wall-sized displays for information visualization applications.
\newblock {\em IEEE Transactions on Visualization and Computer Graphics}, 18(12):2516--2525, 2012. \href{https://doi.org/10.1109/tvcg.2012.251}
{doi: \textsf{%
10\hspace{.1pt}\discretionary{.}{%
}{.}\hspace{.4pt}1109\discretionary{/}{%
}{/}tvcg\hspace{.1pt}\discretionary{.}{%
}{.}\hspace{.4pt}2012\hspace{.1pt}\discretionary{.}{%
}{.}\hspace{.4pt}251}}


\bibitem{bezerianos2013perceptual}
A.~Bezerianos, P.~Isenberg, O.~Chapuis, and W.~Willett.
\newblock Perceptual affordances of wall-sized displays for visualization applications: Color.
\newblock In {\em Proceedings of the CHI Workshop on Interactive, Ultra-High-Resolution Displays (PowerWall)}, 2013.

\bibitem{bi2010effects}
X.~Bi, S.-H. Bae, and R.~Balakrishnan.
\newblock Effects of interior bezels of tiled-monitor large displays on visual search, tunnel steering, and target selection.
\newblock In {\em Proceedings of the SIGCHI conference on Human factors in computing systems}, pp. 65--74, 2010. \href{https://doi.org/10.1145/1753326.1753337}
{doi: \textsf{%
10\hspace{.1pt}\discretionary{.}{%
}{.}\hspace{.4pt}1145\discretionary{/}{%
}{/}1753326\hspace{.1pt}\discretionary{.}{%
}{.}\hspace{.4pt}1753337}}


\bibitem{bowman20043d}
D.~Bowman, E.~Kruijff, J.~J. LaViola~Jr, and I.~P. Poupyrev.
\newblock {\em 3D User interfaces: theory and practice, CourseSmart eTextbook}.
\newblock Addison-Wesley, 2004.

\bibitem{bowman1997travel}
D.~A. Bowman, D.~Koller, and L.~F. Hodges.
\newblock Travel in immersive virtual environments: An evaluation of viewpoint motion control techniques.
\newblock In {\em Proceedings of IEEE 1997 Annual International Symposium on Virtual Reality}, pp. 45--52. IEEE, 1997. \href{https://doi.org/10.1109/vrais.1997.583043}
{doi: \textsf{%
10\hspace{.1pt}\discretionary{.}{%
}{.}\hspace{.4pt}1109\discretionary{/}{%
}{/}vrais\hspace{.1pt}\discretionary{.}{%
}{.}\hspace{.4pt}1997\hspace{.1pt}\discretionary{.}{%
}{.}\hspace{.4pt}583043}}


\bibitem{bradel2013large}
L.~Bradel, A.~Endert, K.~Koch, C.~Andrews, and C.~North.
\newblock Large high resolution displays for co-located collaborative sensemaking: Display usage and territoriality.
\newblock {\em International Journal of Human-Computer Studies}, 71(11):1078--1088, 2013. \href{https://doi.org/10.1016/j.ijhcs.2013.07.004}
{doi: \textsf{%
10\hspace{.1pt}\discretionary{.}{%
}{.}\hspace{.4pt}1016\discretionary{/}{%
}{/}j\hspace{.1pt}\discretionary{.}{%
}{.}\hspace{.4pt}ijhcs\hspace{.1pt}\discretionary{.}{%
}{.}\hspace{.4pt}2013\hspace{.1pt}\discretionary{.}{%
}{.}\hspace{.4pt}07\hspace{.1pt}\discretionary{.}{%
}{.}\hspace{.4pt}004}}


\bibitem{buchsbaum1971neural}
M.~Buchsbaum.
\newblock Neural events and psychophysical law.
\newblock {\em Science}, 172(3982):502--502, 1971. \href{https://doi.org/10.1126/science.172.3982.502}
{doi: \textsf{%
10\hspace{.1pt}\discretionary{.}{%
}{.}\hspace{.4pt}1126\discretionary{/}{%
}{/}science\hspace{.1pt}\discretionary{.}{%
}{.}\hspace{.4pt}172\hspace{.1pt}\discretionary{.}{%
}{.}\hspace{.4pt}3982\hspace{.1pt}\discretionary{.}{%
}{.}\hspace{.4pt}502}}


\bibitem{calmettes2012making}
G.~Calmettes, G.~B. Drummond, and S.~L. Vowler.
\newblock Making do with what we have: use your bootstraps.
\newblock {\em Advances in physiology education}, 36(3):177--180, 2012. \href{https://doi.org/10.1113/expphysiol.2012.068379}
{doi: \textsf{%
10\hspace{.1pt}\discretionary{.}{%
}{.}\hspace{.4pt}1113\discretionary{/}{%
}{/}expphysiol\hspace{.1pt}\discretionary{.}{%
}{.}\hspace{.4pt}2012\hspace{.1pt}\discretionary{.}{%
}{.}\hspace{.4pt}068379}}


\bibitem{chapuis2024comparing}
O.~Chapuis, R.~James, M.~Rafiei, and A.~Bezerianos.
\newblock {Comparing Physical and VR-emulated Ultra-walls in a Classification Task}.
\newblock working paper or preprint, Feb. 2024.

\bibitem{choudhary2021revisiting}
Z.~Choudhary, M.~Gottsacker, K.~Kim, R.~Schubert, J.~Stefanucci, G.~Bruder, and G.~F. Welch.
\newblock Revisiting distance perception with scaled embodied cues in social virtual reality.
\newblock In {\em 2021 IEEE Virtual Reality and 3D User Interfaces (VR)}, pp. 788--797. IEEE, 2021. \href{https://doi.org/10.1109/vr50410.2021.00106}
{doi: \textsf{%
10\hspace{.1pt}\discretionary{.}{%
}{.}\hspace{.4pt}1109\discretionary{/}{%
}{/}vr50410\hspace{.1pt}\discretionary{.}{%
}{.}\hspace{.4pt}2021\hspace{.1pt}\discretionary{.}{%
}{.}\hspace{.4pt}00106}}


\bibitem{cleveland1985elements}
W.~S. Cleveland.
\newblock {\em The elements of graphing data}.
\newblock Wadsworth Publ. Co., 1985. \href{https://doi.org/10.1093/sysbio/34.4.471}
{doi: \textsf{%
10\hspace{.1pt}\discretionary{.}{%
}{.}\hspace{.4pt}1093\discretionary{/}{%
}{/}sysbio\discretionary{/}{%
}{/}34\hspace{.1pt}\discretionary{.}{%
}{.}\hspace{.4pt}4\hspace{.1pt}\discretionary{.}{%
}{.}\hspace{.4pt}471}}


\bibitem{cleveland1984graphical}
W.~S. Cleveland and R.~McGill.
\newblock Graphical perception: Theory, experimentation, and application to the development of graphical methods.
\newblock {\em Journal of the American statistical association}, 79(387):531--554, 1984. \href{https://doi.org/10.2307/2288400}
{doi: \textsf{%
10\hspace{.1pt}\discretionary{.}{%
}{.}\hspace{.4pt}2307\discretionary{/}{%
}{/}2288400}}


\bibitem{cockburn2020threats}
A.~Cockburn, P.~Dragicevic, L.~Besan{\c{c}}on, and C.~Gutwin.
\newblock Threats of a replication crisis in empirical computer science.
\newblock {\em Communications of the ACM}, 63(8):70--79, 2020. \href{https://doi.org/10.1145/3360311}
{doi: \textsf{%
10\hspace{.1pt}\discretionary{.}{%
}{.}\hspace{.4pt}1145\discretionary{/}{%
}{/}3360311}}


\bibitem{czerwinski2003toward}
M.~Czerwinski, G.~Smith, T.~Regan, B.~Meyers, G.~G. Robertson, and G.~K. Starkweather.
\newblock Toward characterizing the productivity benefits of very large displays.
\newblock In {\em Interact}, vol.~3, pp. 9--16, 2003.

\bibitem{dragicevic2016fair}
P.~Dragicevic.
\newblock Fair statistical communication in hci.
\newblock {\em Modern statistical methods for HCI}, pp. 291--330, 2016. \href{https://doi.org/10.1007/978-3-319-26633-6_13}
{doi: \textsf{%
10\hspace{.1pt}\discretionary{.}{%
}{.}\hspace{.4pt}1007\discretionary{/}{%
}{/}978\discretionary{%
}{-}{-}3\discretionary{%
}{-}{-}319\discretionary{%
}{-}{-}26633\discretionary{%
}{-}{-}6\_13}}


\bibitem{dwyer2018immersive}
T.~Dwyer, K.~Marriott, T.~Isenberg, K.~Klein, N.~Riche, F.~Schreiber, W.~Stuerzlinger, and B.~H. Thomas.
\newblock Immersive analytics: An introduction.
\newblock {\em Immersive analytics}, pp. 1--23, 2018. \href{https://doi.org/10.1007/978-3-030-01388-2_1}
{doi: \textsf{%
10\hspace{.1pt}\discretionary{.}{%
}{.}\hspace{.4pt}1007\discretionary{/}{%
}{/}978\discretionary{%
}{-}{-}3\discretionary{%
}{-}{-}030\discretionary{%
}{-}{-}01388\discretionary{%
}{-}{-}2\_1}}


\bibitem{el2019survey}
F.~El~Jamiy and R.~Marsh.
\newblock Survey on depth perception in head mounted displays: distance estimation in virtual reality, augmented reality, and mixed reality.
\newblock {\em IET Image Processing}, 13(5):707--712, 2019. \href{https://doi.org/10.1049/iet-ipr.2018.5920}
{doi: \textsf{%
10\hspace{.1pt}\discretionary{.}{%
}{.}\hspace{.4pt}1049\discretionary{/}{%
}{/}iet\discretionary{%
}{-}{-}ipr\hspace{.1pt}\discretionary{.}{%
}{.}\hspace{.4pt}2018\hspace{.1pt}\discretionary{.}{%
}{.}\hspace{.4pt}5920}}


\bibitem{endert2011visual}
A.~Endert, C.~Andrews, Y.-H. Lee, and C.~North.
\newblock Visual encodings that support physical navigation on large displays.
\newblock In {\em Graphics Interface}, pp. 103--110, 2011.

\bibitem{ens2014ethereal}
B.~Ens, J.~D. Hincapi{\'e}-Ramos, and P.~Irani.
\newblock Ethereal planes: a design framework for 2d information space in 3d mixed reality environments.
\newblock In {\em Proceedings of the 2nd ACM symposium on Spatial user interaction}, pp. 2--12, 2014.

\bibitem{erickson2020effects}
A.~Erickson, K.~Kim, G.~Bruder, and G.~F. Welch.
\newblock Effects of dark mode graphics on visual acuity and fatigue with virtual reality head-mounted displays.
\newblock In {\em 2020 IEEE Conference on virtual reality and 3D user interfaces (VR)}, pp. 434--442. IEEE, 2020.

\bibitem{etemadpour2013effect}
R.~Etemadpour, E.~Monson, and L.~Linsen.
\newblock The effect of stereoscopic immersive environments on projection-based multi-dimensional data visualization.
\newblock In {\em 2013 17th International Conference on Information Visualisation}, pp. 389--397. IEEE, 2013.

\bibitem{han2023evaluating}
D.~Han and I.~Cho.
\newblock Evaluating 3d user interaction techniques on spatial working memory for 3d scatter plot exploration in immersive analytics.
\newblock In {\em 2023 IEEE International Symposium on Mixed and Augmented Reality (ISMAR)}, pp. 513--522. IEEE, 2023. \href{https://doi.org/10.1109/ismar59233.2023.00066}
{doi: \textsf{%
10\hspace{.1pt}\discretionary{.}{%
}{.}\hspace{.4pt}1109\discretionary{/}{%
}{/}ismar59233\hspace{.1pt}\discretionary{.}{%
}{.}\hspace{.4pt}2023\hspace{.1pt}\discretionary{.}{%
}{.}\hspace{.4pt}00066}}


\bibitem{han2022portal}
D.~Han, D.~Kim, and I.~Cho.
\newblock Portal: Portal widget for remote target acquisition and control in immersive virtual environments.
\newblock In {\em Proceedings of the 28th ACM Symposium on Virtual Reality Software and Technology}, pp. 1--11, 2022. \href{https://doi.org/10.1145/3562939.3565639}
{doi: \textsf{%
10\hspace{.1pt}\discretionary{.}{%
}{.}\hspace{.4pt}1145\discretionary{/}{%
}{/}3562939\hspace{.1pt}\discretionary{.}{%
}{.}\hspace{.4pt}3565639}}


\bibitem{henry1993spatial}
D.~Henry and T.~Furness.
\newblock Spatial perception in virtual environments: Evaluating an architectural application.
\newblock In {\em Proceedings of IEEE virtual reality annual international symposium}, pp. 33--40. IEEE, 1993. \href{https://doi.org/10.1109/vrais.1993.380801}
{doi: \textsf{%
10\hspace{.1pt}\discretionary{.}{%
}{.}\hspace{.4pt}1109\discretionary{/}{%
}{/}vrais\hspace{.1pt}\discretionary{.}{%
}{.}\hspace{.4pt}1993\hspace{.1pt}\discretionary{.}{%
}{.}\hspace{.4pt}380801}}


\bibitem{higgins2004introduction}
J.~J. Higgins.
\newblock {\em An introduction to modern nonparametric statistics}.
\newblock Brooks/Cole Pacific Grove, CA, 2004.

\bibitem{in2024evaluating}
S.~In, E.~Krokos, K.~Whitley, C.~North, and Y.~Yang.
\newblock Evaluating navigation and comparison performance of computational notebooks on desktop and in virtual reality.
\newblock In {\em Proceedings of the CHI Conference on Human Factors in Computing Systems}, pp. 1--15, 2024.

\bibitem{EyeMeasure}
B.~Inc.
\newblock Eyemeasure, 2024.

\bibitem{interrante2006distance}
V.~Interrante, B.~Ries, and L.~Anderson.
\newblock Distance perception in immersive virtual environments, revisited.
\newblock In {\em IEEE virtual reality conference (VR 2006)}, pp. 3--10. IEEE, 2006. \href{https://doi.org/10.1109/vr.2006.52}
{doi: \textsf{%
10\hspace{.1pt}\discretionary{.}{%
}{.}\hspace{.4pt}1109\discretionary{/}{%
}{/}vr\hspace{.1pt}\discretionary{.}{%
}{.}\hspace{.4pt}2006\hspace{.1pt}\discretionary{.}{%
}{.}\hspace{.4pt}52}}


\bibitem{isenberg2011co}
P.~Isenberg, D.~Fisher, S.~A. Paul, M.~R. Morris, K.~Inkpen, and M.~Czerwinski.
\newblock Co-located collaborative visual analytics around a tabletop display.
\newblock {\em IEEE Transactions on visualization and Computer Graphics}, 18(5):689--702, 2011. \href{https://doi.org/10.1109/tvcg.2011.287}
{doi: \textsf{%
10\hspace{.1pt}\discretionary{.}{%
}{.}\hspace{.4pt}1109\discretionary{/}{%
}{/}tvcg\hspace{.1pt}\discretionary{.}{%
}{.}\hspace{.4pt}2011\hspace{.1pt}\discretionary{.}{%
}{.}\hspace{.4pt}287}}


\bibitem{jakobsen2014up}
M.~R. Jakobsen and K.~Hornb{\ae}k.
\newblock Up close and personal: Collaborative work on a high-resolution multitouch wall display.
\newblock {\em ACM Transactions on Computer-Human Interaction (TOCHI)}, 21(2):1--34, 2014. \href{https://doi.org/10.1145/2576099}
{doi: \textsf{%
10\hspace{.1pt}\discretionary{.}{%
}{.}\hspace{.4pt}1145\discretionary{/}{%
}{/}2576099}}


\bibitem{James2023evaluating}
R.~James, A.~Bezerianos, and O.~Chapuis.
\newblock Evaluating the extension of wall displays with ar for collaborative work.
\newblock In {\em Proceedings of the 2023 CHI Conference on Human Factors in Computing Systems}, CHI '23,  article no. 99,  17 pages. Association for Computing Machinery, New York, NY, USA, 2023. \href{https://doi.org/10.1145/3544548.3580752}
{doi: \textsf{%
10\hspace{.1pt}\discretionary{.}{%
}{.}\hspace{.4pt}1145\discretionary{/}{%
}{/}3544548\hspace{.1pt}\discretionary{.}{%
}{.}\hspace{.4pt}3580752}}


\bibitem{kiluk2023impact}
A.~Kiluk, V.~Paneva, S.~Seinfeld, and J.~M{\"u}ller.
\newblock The impact of different virtual work environments on flow, performance, user emotions, and preferences.
\newblock In {\em 2023 IEEE International Symposium on Mixed and Augmented Reality Adjunct (ISMAR-Adjunct)}, pp. 276--282. IEEE, 2023.

\bibitem{kim2018virtual}
H.~K. Kim, J.~Park, Y.~Choi, and M.~Choe.
\newblock Virtual reality sickness questionnaire (vrsq): Motion sickness measurement index in a virtual reality environment.
\newblock {\em Applied ergonomics}, 69:66--73, 2018. \href{https://doi.org/10.1016/j.apergo.2017.12.016}
{doi: \textsf{%
10\hspace{.1pt}\discretionary{.}{%
}{.}\hspace{.4pt}1016\discretionary{/}{%
}{/}j\hspace{.1pt}\discretionary{.}{%
}{.}\hspace{.4pt}apergo\hspace{.1pt}\discretionary{.}{%
}{.}\hspace{.4pt}2017\hspace{.1pt}\discretionary{.}{%
}{.}\hspace{.4pt}12\hspace{.1pt}\discretionary{.}{%
}{.}\hspace{.4pt}016}}


\bibitem{kobayashi2021translating}
D.~Kobayashi, N.~Kirshenbaum, R.~S. Tabalba, R.~Theriot, and J.~Leigh.
\newblock Translating the benefits of wide-band display environments into an {XR} space.
\newblock In {\em Proceedings of the Symposium on Spatial User Interaction (SUI)}, pp. 9:1--9:11, 2021. \href{https://doi.org/10.1145/3485279.3485294}
{doi: \textsf{%
10\hspace{.1pt}\discretionary{.}{%
}{.}\hspace{.4pt}1145\discretionary{/}{%
}{/}3485279\hspace{.1pt}\discretionary{.}{%
}{.}\hspace{.4pt}3485294}}


\bibitem{kraus2022immersive}
M.~Kraus, J.~Fuchs, B.~Sommer, K.~Klein, U.~Engelke, D.~Keim, and F.~Schreiber.
\newblock Immersive analytics with abstract 3d visualizations: A survey.
\newblock In {\em Computer Graphics Forum}, vol.~41, pp. 201--229. Wiley Online Library, 2022. \href{https://doi.org/10.1111/cgf.14430}
{doi: \textsf{%
10\hspace{.1pt}\discretionary{.}{%
}{.}\hspace{.4pt}1111\discretionary{/}{%
}{/}cgf\hspace{.1pt}\discretionary{.}{%
}{.}\hspace{.4pt}14430}}


\bibitem{kraus2019impact}
M.~Kraus, N.~Weiler, D.~Oelke, J.~Kehrer, D.~A. Keim, and J.~Fuchs.
\newblock The impact of immersion on cluster identification tasks.
\newblock {\em IEEE Transactions on Visualization and Computer Graphics}, 26(1):525--535, 2019. \href{https://doi.org/10.1109/tvcg.2019.2934395}
{doi: \textsf{%
10\hspace{.1pt}\discretionary{.}{%
}{.}\hspace{.4pt}1109\discretionary{/}{%
}{/}tvcg\hspace{.1pt}\discretionary{.}{%
}{.}\hspace{.4pt}2019\hspace{.1pt}\discretionary{.}{%
}{.}\hspace{.4pt}2934395}}


\bibitem{laviola20173d}
J.~J. LaViola~Jr, E.~Kruijff, R.~P. McMahan, D.~Bowman, and I.~P. Poupyrev.
\newblock {\em 3D user interfaces: theory and practice}.
\newblock Addison-Wesley Professional, 2017.

\bibitem{leigh2019usage}
J.~Leigh, D.~Kobayashi, N.~Kirshenbaum, T.~Wooton, A.~Gonzalez, L.~Renambot, A.~Johnson, M.~Brown, A.~Burks, K.~Bharadwaj, et~al.
\newblock Usage patterns of wideband display environments in e-science research, development and training.
\newblock In {\em 2019 15th International Conference on eScience (eScience)}, pp. 301--310. IEEE, 2019. \href{https://doi.org/10.1109/escience.2019.00041}
{doi: \textsf{%
10\hspace{.1pt}\discretionary{.}{%
}{.}\hspace{.4pt}1109\discretionary{/}{%
}{/}escience\hspace{.1pt}\discretionary{.}{%
}{.}\hspace{.4pt}2019\hspace{.1pt}\discretionary{.}{%
}{.}\hspace{.4pt}00041}}


\bibitem{levine1999pace}
R.~V. Levine and A.~Norenzayan.
\newblock The pace of life in 31 countries.
\newblock {\em Journal of cross-cultural psychology}, 30(2):178--205, 1999. \href{https://doi.org/10.1177/0022022199030002003}
{doi: \textsf{%
10\hspace{.1pt}\discretionary{.}{%
}{.}\hspace{.4pt}1177\discretionary{/}{%
}{/}0022022199030002003}}


\bibitem{leyrer2015eye}
M.~Leyrer, S.~A. Linkenauger, H.~H. B{\"u}lthoff, and B.~J. Mohler.
\newblock Eye height manipulations: A possible solution to reduce underestimation of egocentric distances in head-mounted displays.
\newblock {\em ACM Transactions on Applied Perception (TAP)}, 12(1):1--23, 2015. \href{https://doi.org/10.1145/2699254}
{doi: \textsf{%
10\hspace{.1pt}\discretionary{.}{%
}{.}\hspace{.4pt}1145\discretionary{/}{%
}{/}2699254}}


\bibitem{lisle2021sensemaking}
L.~Lisle, K.~Davidson, E.~J. Gitre, C.~North, and D.~A. Bowman.
\newblock Sensemaking strategies with immersive space to think.
\newblock In {\em 2021 IEEE Virtual Reality and 3D User Interfaces (VR)}, pp. 529--537. IEEE, 2021. \href{https://doi.org/10.1109/vr50410.2021.00077}
{doi: \textsf{%
10\hspace{.1pt}\discretionary{.}{%
}{.}\hspace{.4pt}1109\discretionary{/}{%
}{/}vr50410\hspace{.1pt}\discretionary{.}{%
}{.}\hspace{.4pt}2021\hspace{.1pt}\discretionary{.}{%
}{.}\hspace{.4pt}00077}}


\bibitem{liu2014effects}
C.~Liu, O.~Chapuis, M.~Beaudouin-Lafon, E.~Lecolinet, and W.~E. Mackay.
\newblock Effects of display size and navigation type on a classification task.
\newblock In {\em Proceedings of the Conference on Human Factors in Computing Systems}, pp. 4147--4156, 2014. \href{https://doi.org/10.1145/2556288.2557020}
{doi: \textsf{%
10\hspace{.1pt}\discretionary{.}{%
}{.}\hspace{.4pt}1145\discretionary{/}{%
}{/}2556288\hspace{.1pt}\discretionary{.}{%
}{.}\hspace{.4pt}2557020}}


\bibitem{liu2023datadancing}
J.~Liu, B.~Ens, A.~Prouzeau, J.~Smiley, I.~K. Nixon, S.~Goodwin, and T.~Dwyer.
\newblock {DataDancing}: An exploration of the design space for visualisation view management for 3d surfaces and spaces.
\newblock In {\em Proceedings of the Conference on Human Factors in Computing Systems (CHI)}, pp. 379:1--379:17, 2023. \href{https://doi.org/10.1145/3544548.3580827}
{doi: \textsf{%
10\hspace{.1pt}\discretionary{.}{%
}{.}\hspace{.4pt}1145\discretionary{/}{%
}{/}3544548\hspace{.1pt}\discretionary{.}{%
}{.}\hspace{.4pt}3580827}}


\bibitem{liu2020design}
J.~Liu, A.~Prouzeau, B.~Ens, and T.~Dwyer.
\newblock Design and evaluation of interactive small multiples data visualisation in immersive spaces.
\newblock In {\em Proceedings of the Conference on Virtual Reality and 3D User Interfaces (VR)}, pp. 588--597. IEEE, 2020. \href{https://doi.org/10.1109/VR46266.2020.00081}
{doi: \textsf{%
10\hspace{.1pt}\discretionary{.}{%
}{.}\hspace{.4pt}1109\discretionary{/}{%
}{/}VR46266\hspace{.1pt}\discretionary{.}{%
}{.}\hspace{.4pt}2020\hspace{.1pt}\discretionary{.}{%
}{.}\hspace{.4pt}00081}}


\bibitem{liu2022effects}
J.~Liu, A.~Prouzeau, B.~Ens, and T.~Dwyer.
\newblock Effects of display layout on spatial memory for immersive environments.
\newblock {\em Proceedings of the ACM on Human-Computer Interaction}, 6(ISS):468--488, 2022. \href{https://doi.org/10.1145/3567729}
{doi: \textsf{%
10\hspace{.1pt}\discretionary{.}{%
}{.}\hspace{.4pt}1145\discretionary{/}{%
}{/}3567729}}


\bibitem{loomis2003visual}
J.~M. Loomis and J.~M. Knapp.
\newblock Visual perception of egocentric distance in real and virtual environments.
\newblock In {\em Virtual and adaptive environments}, pp. 21--46. CRC Press, 2003. \href{https://doi.org/10.1201/9781410608888.pt1}
{doi: \textsf{%
10\hspace{.1pt}\discretionary{.}{%
}{.}\hspace{.4pt}1201\discretionary{/}{%
}{/}9781410608888\hspace{.1pt}\discretionary{.}{%
}{.}\hspace{.4pt}pt1}}


\bibitem{luo2022should}
W.~Luo, A.~Lehmann, H.~Widengren, and R.~Dachselt.
\newblock Where should we put it? layout and placement strategies of documents in augmented reality for collaborative sensemaking.
\newblock In {\em Proceedings of the 2022 CHI Conference on Human Factors in Computing Systems}, pp. 1--16, 2022. \href{https://doi.org/10.1145/3491102.3501946}
{doi: \textsf{%
10\hspace{.1pt}\discretionary{.}{%
}{.}\hspace{.4pt}1145\discretionary{/}{%
}{/}3491102\hspace{.1pt}\discretionary{.}{%
}{.}\hspace{.4pt}3501946}}


\bibitem{mackinlay1985expressiveness}
J.~Mackinlay and M.~R. Genesereth.
\newblock Expressiveness and language choice.
\newblock {\em Data \& Knowledge Engineering}, 1(1):17--29, 1985. \href{https://doi.org/10.1016/0169-023x(85)90025-4}
{doi: \textsf{%
10\hspace{.1pt}\discretionary{.}{%
}{.}\hspace{.4pt}1016\discretionary{/}{%
}{/}0169\discretionary{%
}{-}{-}023x\discretionary{%
}{(}{(}85\discretionary{)}{%
}{)}90025\discretionary{%
}{-}{-}4}}


\bibitem{mohler2010effect}
B.~J. Mohler, S.~H. Creem-Regehr, W.~B. Thompson, and H.~H. B{\"u}lthoff.
\newblock The effect of viewing a self-avatar on distance judgments in an hmd-based virtual environment.
\newblock {\em Presence}, 19(3):230--242, 2010. \href{https://doi.org/10.1162/pres.19.3.230}
{doi: \textsf{%
10\hspace{.1pt}\discretionary{.}{%
}{.}\hspace{.4pt}1162\discretionary{/}{%
}{/}pres\hspace{.1pt}\discretionary{.}{%
}{.}\hspace{.4pt}19\hspace{.1pt}\discretionary{.}{%
}{.}\hspace{.4pt}3\hspace{.1pt}\discretionary{.}{%
}{.}\hspace{.4pt}230}}


\bibitem{mohler2007visual}
B.~J. Mohler, W.~B. Thompson, S.~H. Creem-Regehr, H.~L. Pick, and W.~H. Warren.
\newblock Visual flow influences gait transition speed and preferred walking speed.
\newblock {\em Experimental brain research}, 181:221--228, 2007. \href{https://doi.org/10.1007/s00221-007-0917-0}
{doi: \textsf{%
10\hspace{.1pt}\discretionary{.}{%
}{.}\hspace{.4pt}1007\discretionary{/}{%
}{/}s00221\discretionary{%
}{-}{-}007\discretionary{%
}{-}{-}0917\discretionary{%
}{-}{-}0}}


\bibitem{munzner2014visualization}
T.~Munzner.
\newblock {\em Visualization analysis and design}.
\newblock CRC press, 2014. \href{https://doi.org/10.1201/b17511}
{doi: \textsf{%
10\hspace{.1pt}\discretionary{.}{%
}{.}\hspace{.4pt}1201\discretionary{/}{%
}{/}b17511}}


\bibitem{phillips2010avatar}
L.~Phillips, B.~Ries, M.~Kaeding, and V.~Interrante.
\newblock Avatar self-embodiment enhances distance perception accuracy in non-photorealistic immersive virtual environments.
\newblock In {\em 2010 IEEE virtual reality conference (VR)}, pp. 115--1148. IEEE, 2010. \href{https://doi.org/10.1109/vr.2010.5444802}
{doi: \textsf{%
10\hspace{.1pt}\discretionary{.}{%
}{.}\hspace{.4pt}1109\discretionary{/}{%
}{/}vr\hspace{.1pt}\discretionary{.}{%
}{.}\hspace{.4pt}2010\hspace{.1pt}\discretionary{.}{%
}{.}\hspace{.4pt}5444802}}


\bibitem{prouzeau2017tradeoffs}
A.~Prouzeau, A.~Bezerianos, and O.~Chapuis.
\newblock Trade-offs between a vertical shared display and two desktops in a collaborative path-finding task.
\newblock In {\em Proceedings of Graphics Interface}, GI '17,  6 pages. CHCCS, May 2017. \href{https://doi.org/10.5555/3141475.3141516}
{doi: \textsf{%
10\hspace{.1pt}\discretionary{.}{%
}{.}\hspace{.4pt}5555\discretionary{/}{%
}{/}3141475\hspace{.1pt}\discretionary{.}{%
}{.}\hspace{.4pt}3141516}}


\bibitem{reipschlager2020personal}
P.~Reipschlager, T.~Flemisch, and R.~Dachselt.
\newblock Personal augmented reality for information visualization on large interactive displays.
\newblock {\em IEEE Transactions on Visualization and Computer Graphics}, 27(2):1182--1192, 2020. \href{https://doi.org/10.1109/tvcg.2020.3030460}
{doi: \textsf{%
10\hspace{.1pt}\discretionary{.}{%
}{.}\hspace{.4pt}1109\discretionary{/}{%
}{/}tvcg\hspace{.1pt}\discretionary{.}{%
}{.}\hspace{.4pt}2020\hspace{.1pt}\discretionary{.}{%
}{.}\hspace{.4pt}3030460}}


\bibitem{ries2008effect}
B.~Ries, V.~Interrante, M.~Kaeding, and L.~Anderson.
\newblock The effect of self-embodiment on distance perception in immersive virtual environments.
\newblock In {\em Proceedings of the 2008 ACM symposium on Virtual reality software and technology}, pp. 167--170, 2008. \href{https://doi.org/10.1145/1450579.1450614}
{doi: \textsf{%
10\hspace{.1pt}\discretionary{.}{%
}{.}\hspace{.4pt}1145\discretionary{/}{%
}{/}1450579\hspace{.1pt}\discretionary{.}{%
}{.}\hspace{.4pt}1450614}}


\bibitem{ries2009analyzing}
B.~Ries, V.~Interrante, M.~Kaeding, and L.~Phillips.
\newblock Analyzing the effect of a virtual avatar's geometric and motion fidelity on ego-centric spatial perception in immersive virtual environments.
\newblock In {\em Proceedings of the 16th ACM symposium on virtual reality software and technology}, pp. 59--66, 2009. \href{https://doi.org/10.1145/1643928.1643943}
{doi: \textsf{%
10\hspace{.1pt}\discretionary{.}{%
}{.}\hspace{.4pt}1145\discretionary{/}{%
}{/}1643928\hspace{.1pt}\discretionary{.}{%
}{.}\hspace{.4pt}1643943}}


\bibitem{ruddle2009benefits}
R.~A. Ruddle and S.~Lessels.
\newblock The benefits of using a walking interface to navigate virtual environments.
\newblock {\em ACM Transactions on Computer-Human Interaction (TOCHI)}, 16(1):1--18, 2009. \href{https://doi.org/10.1145/1502800.1502805}
{doi: \textsf{%
10\hspace{.1pt}\discretionary{.}{%
}{.}\hspace{.4pt}1145\discretionary{/}{%
}{/}1502800\hspace{.1pt}\discretionary{.}{%
}{.}\hspace{.4pt}1502805}}


\bibitem{satkowski2022above}
M.~Satkowski, R.~Rzayev, E.~Goebel, and R.~Dachselt.
\newblock Above \& below: investigating ceiling and floor for augmented reality content placement.
\newblock In {\em 2022 IEEE International Symposium on Mixed and Augmented Reality (ISMAR)}, pp. 518--527. IEEE, 2022. \href{https://doi.org/10.1109/ismar55827.2022.00068}
{doi: \textsf{%
10\hspace{.1pt}\discretionary{.}{%
}{.}\hspace{.4pt}1109\discretionary{/}{%
}{/}ismar55827\hspace{.1pt}\discretionary{.}{%
}{.}\hspace{.4pt}2022\hspace{.1pt}\discretionary{.}{%
}{.}\hspace{.4pt}00068}}


\bibitem{satriadi2020maps}
K.~A. Satriadi, B.~Ens, M.~Cordeil, T.~Czauderna, and B.~Jenny.
\newblock Maps around me: 3d multiview layouts in immersive spaces.
\newblock {\em Proceedings of the ACM on Human-Computer Interaction}, 4(ISS):1--20, 2020. \href{https://doi.org/10.1145/3427329}
{doi: \textsf{%
10\hspace{.1pt}\discretionary{.}{%
}{.}\hspace{.4pt}1145\discretionary{/}{%
}{/}3427329}}


\bibitem{shupp2009shaping}
L.~Shupp, C.~Andrews, M.~Dickey-Kurdziolek, B.~Yost, and C.~North.
\newblock Shaping the display of the future: The effects of display size and curvature on user performance and insights.
\newblock {\em Human--Computer Interaction}, 24(1-2):230--272, 2009. \href{https://doi.org/10.1080/07370020902739429}
{doi: \textsf{%
10\hspace{.1pt}\discretionary{.}{%
}{.}\hspace{.4pt}1080\discretionary{/}{%
}{/}07370020902739429}}


\bibitem{shupp2006evaluation}
L.~Shupp, R.~Ball, B.~Yost, J.~Booker, and C.~North.
\newblock Evaluation of viewport size and curvature of large, high-resolution displays.
\newblock In {\em Graphics Interface}, pp. 123--130, 2006.

\bibitem{tan2003similar}
D.~S. Tan, D.~Gergle, P.~Scupelli, and R.~Pausch.
\newblock With similar visual angles, larger displays improve spatial performance.
\newblock In {\em Proceedings of the SIGCHI conference on Human factors in computing systems}, pp. 217--224, 2003. \href{https://doi.org/10.1145/642611.642650}
{doi: \textsf{%
10\hspace{.1pt}\discretionary{.}{%
}{.}\hspace{.4pt}1145\discretionary{/}{%
}{/}642611\hspace{.1pt}\discretionary{.}{%
}{.}\hspace{.4pt}642650}}


\bibitem{thompson2011visual}
W.~Thompson, R.~Fleming, S.~Creem-Regehr, and J.~K. Stefanucci.
\newblock {\em Visual perception from a computer graphics perspective}.
\newblock CRC press, 2011.

\bibitem{wagner2018immersive}
J.~A. Wagner~Filho, M.~F. Rey, C.~M. Freitas, and L.~Nedel.
\newblock Immersive visualization of abstract information: An evaluation on dimensionally-reduced data scatterplots.
\newblock In {\em 2018 IEEE Conference on Virtual Reality and 3D User Interfaces (VR)}, pp. 483--490. IEEE, 2018. \href{https://doi.org/10.1109/vr.2018.8447558}
{doi: \textsf{%
10\hspace{.1pt}\discretionary{.}{%
}{.}\hspace{.4pt}1109\discretionary{/}{%
}{/}vr\hspace{.1pt}\discretionary{.}{%
}{.}\hspace{.4pt}2018\hspace{.1pt}\discretionary{.}{%
}{.}\hspace{.4pt}8447558}}


\bibitem{wallace2014effect}
J.~R. Wallace, D.~Vogel, and E.~Lank.
\newblock Effect of bezel presence and width on visual search.
\newblock In {\em Proceedings of The International Symposium on Pervasive Displays}, pp. 118--123, 2014. \href{https://doi.org/10.1145/2611009.2611019}
{doi: \textsf{%
10\hspace{.1pt}\discretionary{.}{%
}{.}\hspace{.4pt}1145\discretionary{/}{%
}{/}2611009\hspace{.1pt}\discretionary{.}{%
}{.}\hspace{.4pt}2611019}}


\bibitem{ware2010visual}
C.~Ware.
\newblock {\em Visual thinking for design}.
\newblock Elsevier, 2010.

\bibitem{wigdor2006effects}
D.~Wigdor, C.~Shen, C.~Forlines, and R.~Balakrishnan.
\newblock Effects of display position and control space orientation on user preference and performance.
\newblock In {\em Proceedings of the SIGCHI conference on human factors in computing systems}, pp. 309--318, 2006. \href{https://doi.org/10.1145/1124772.1124819}
{doi: \textsf{%
10\hspace{.1pt}\discretionary{.}{%
}{.}\hspace{.4pt}1145\discretionary{/}{%
}{/}1124772\hspace{.1pt}\discretionary{.}{%
}{.}\hspace{.4pt}1124819}}


\bibitem{wigdor2007perception}
D.~Wigdor, C.~Shen, C.~Forlines, and R.~Balakrishnan.
\newblock Perception of elementary graphical elements in tabletop and multi-surface environments.
\newblock In {\em Proceedings of the SIGCHI conference on Human factors in computing systems}, pp. 473--482, 2007. \href{https://doi.org/10.1145/1240624.1240701}
{doi: \textsf{%
10\hspace{.1pt}\discretionary{.}{%
}{.}\hspace{.4pt}1145\discretionary{/}{%
}{/}1240624\hspace{.1pt}\discretionary{.}{%
}{.}\hspace{.4pt}1240701}}


\bibitem{yost2007beyond}
B.~Yost, Y.~Haciahmetoglu, and C.~North.
\newblock Beyond visual acuity: the perceptual scalability of information visualizations for large displays.
\newblock In {\em Proceedings of the SIGCHI conference on Human factors in computing systems}, pp. 101--110, 2007. \href{https://doi.org/10.1145/1240624.1240639}
{doi: \textsf{%
10\hspace{.1pt}\discretionary{.}{%
}{.}\hspace{.4pt}1145\discretionary{/}{%
}{/}1240624\hspace{.1pt}\discretionary{.}{%
}{.}\hspace{.4pt}1240639}}


\end{thebibliography}

\clearpage

\appendix 
  \twocolumn[\centering
  {\textsf{\huge Perception of Visual Variables on Virtual Wall-Sized Tiled Displays in Immersive Environments  }}
  \vskip 5pt
  \large \textsf{Appendix}
  \vskip 10pt
\raggedright In this appendix we provide additional images that show analysis results beyond the material that we could include in the main paper due to space limitations.\vspace{.5cm}
]


\renewcommand{\figurename}{Appendix Fig.}
 \renewcommand\thefigure{\arabic{figure}}  
 \setcounter{figure}{0} 
 
\section{VR Device}\label{appendix_vr_device}

\begin{figure}[h]
\centering
\includegraphics[width=.9\columnwidth]{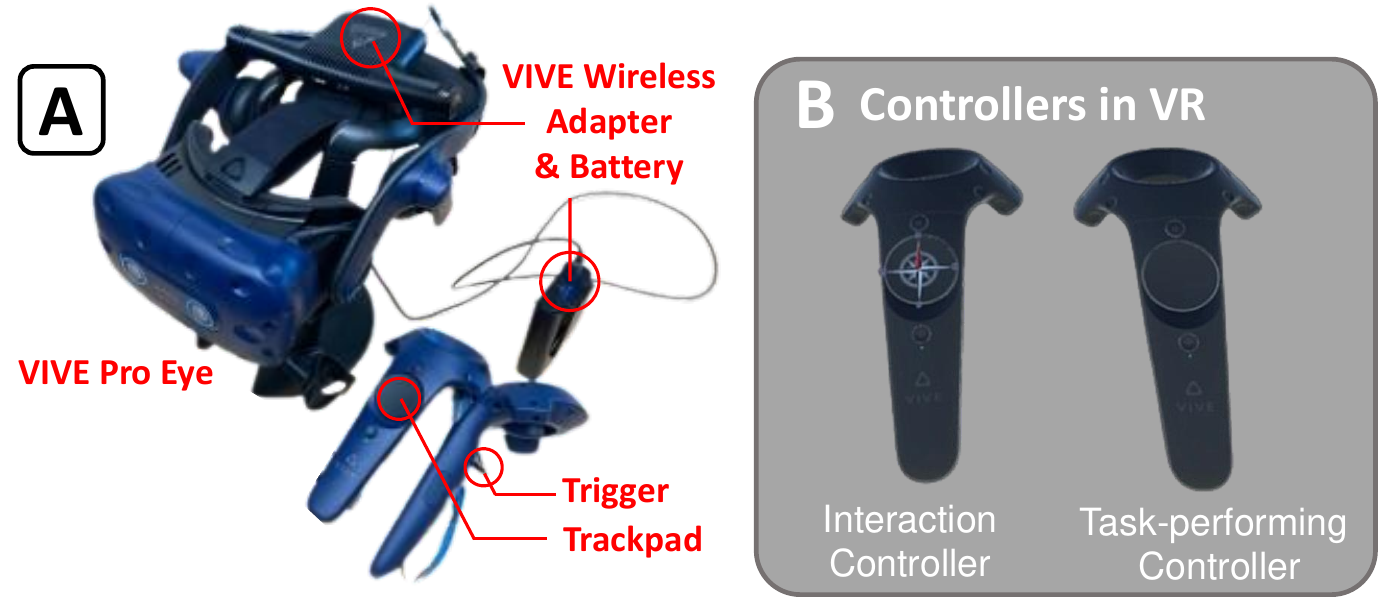}
\caption{
 (A) VR devices used for the user studies including the Vive Pro Eye, two controllers, and wireless adapter, and a battery. (B) Two controllers were seen by participants within the VR space. They are called the task-performing controller and interaction controller. Participants are instructed to hold the task-performing controller in their dominant hand and the interaction controller in their non-dominant hand.}
\end{figure}

\vspace{-.5cm}
\begin{figure}[h]
\section{Additional Result From Study 1}\label{appendix_study1_results}
\centering
\includegraphics[width=.78\columnwidth]{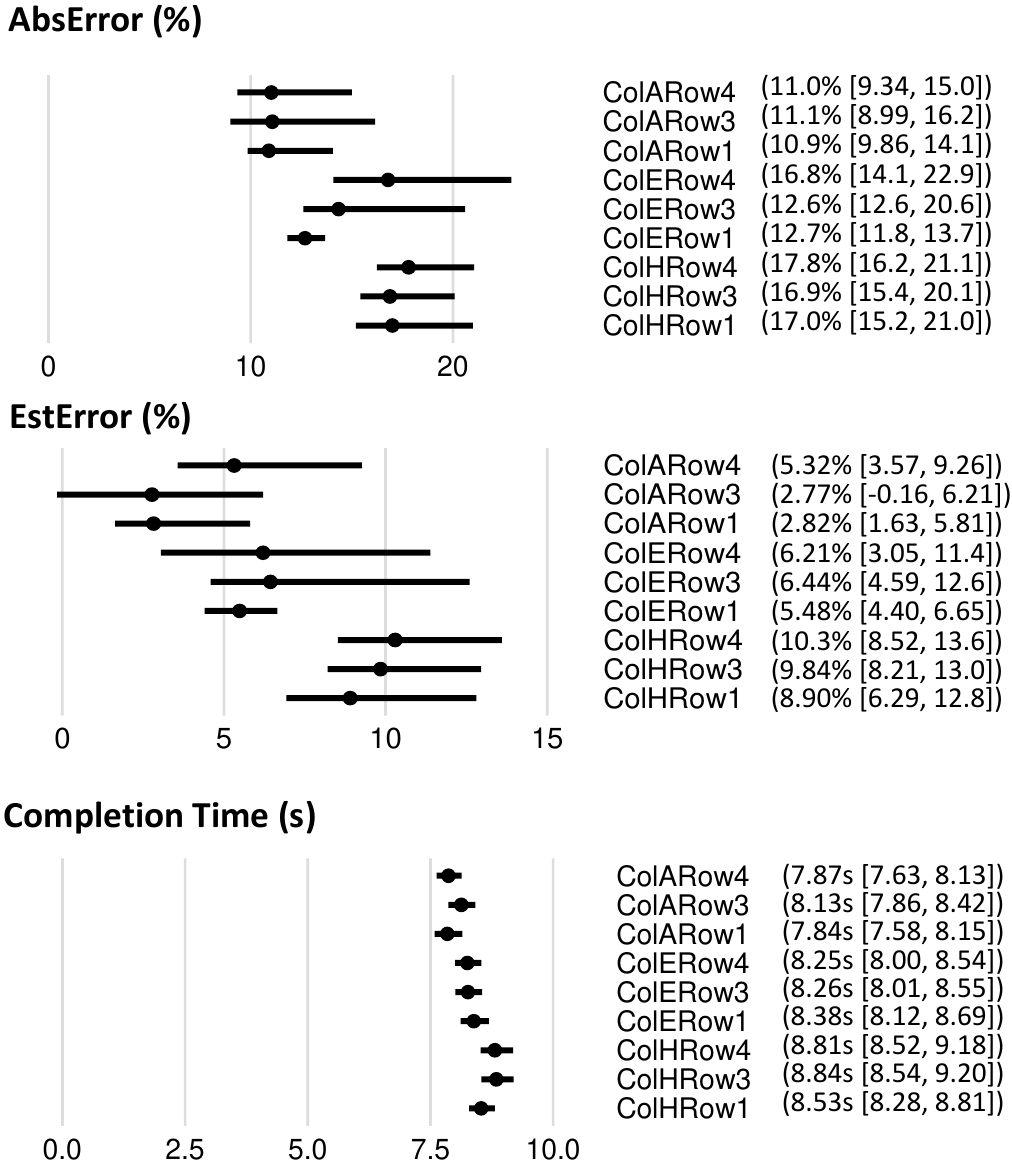}\label{appendix_fig_2}
\caption{Study 1 Results: Detailed Results by Modulus Locations}
\end{figure}

\begin{figure}[h!]
\centering
\includegraphics[width=.78\columnwidth]{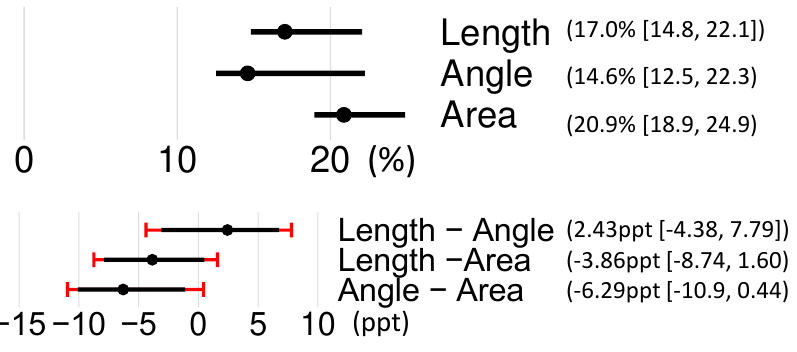}
\caption{Study 1 Results: AbsErr by \studyFactors{Visual Variable} in the \studyFactors{Flat} condition}\label{s1_flat_visVariable_error}
\end{figure}

\begin{figure}[h!]
\centering
\includegraphics[width=.78\columnwidth]{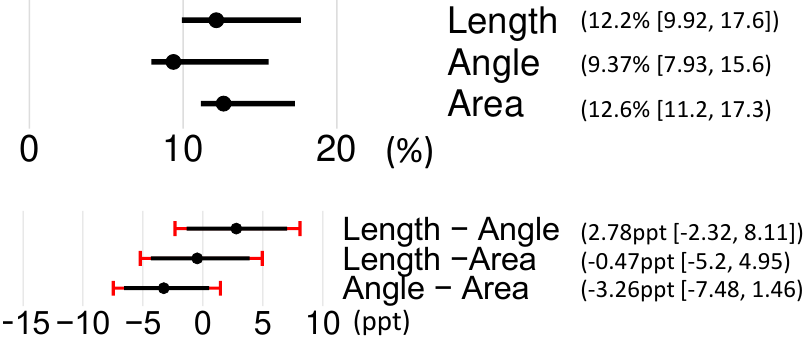}
\caption{Study 1 Results: AbsErr by \studyFactors{Visual Variable} in the \studyFactors{Cylinder} condition}\label{s1_cylinder_visVariable_error}
\includegraphics[width=.78\columnwidth]{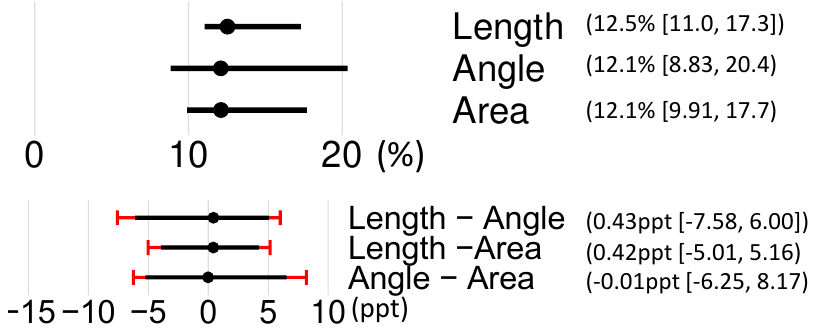}
\caption{Study 1 Results: AbsErr by \studyFactors{Visual Variable} in the \studyFactors{Cockpit} condition}\label{s1_cockpit_visVariable_error}
\end{figure}

\begin{figure}[h]
\section{Additional Result From Study 2}\label{appendix_study2_results}

\includegraphics[width=1\columnwidth]{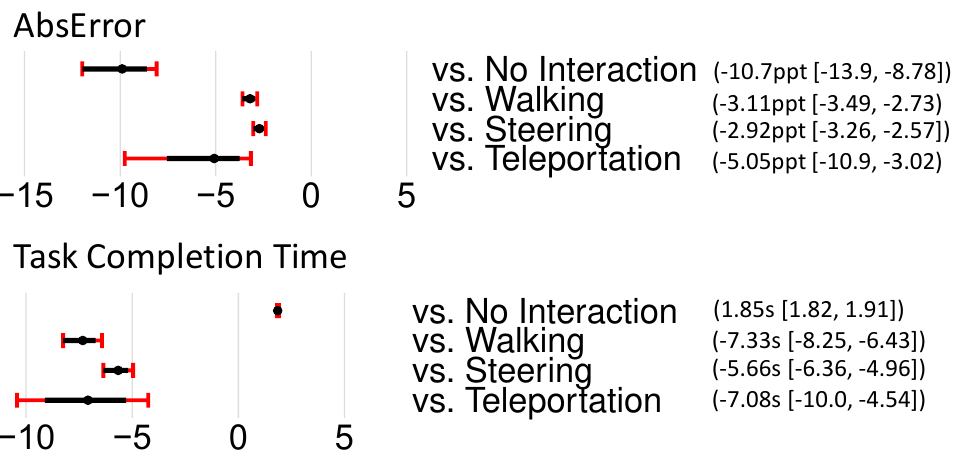}\label{appendix_fig_3}
\caption{Study 2 Results: pairwise comparison results between \studyFactors{Selection} with \studyFactors{Personal} and interaction techniques with \studyFactors{Frontal}}
\includegraphics[width=1\columnwidth]{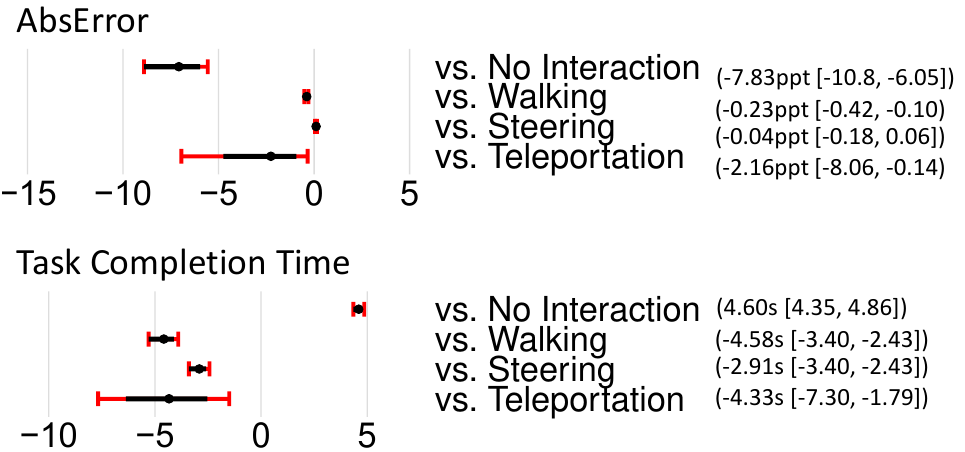}\label{appendix_fig_3}
\caption{Study 2 Results: pairwise comparison results between \studyFactors{Walking} with \studyFactors{Personal} and interaction techniques with \studyFactors{Frontal}}
\end{figure}

\makeatletter
\setlength{\@fptop}{0pt}
\makeatother

\begin{figure}[p]
\centering
\includegraphics[width=1\columnwidth]{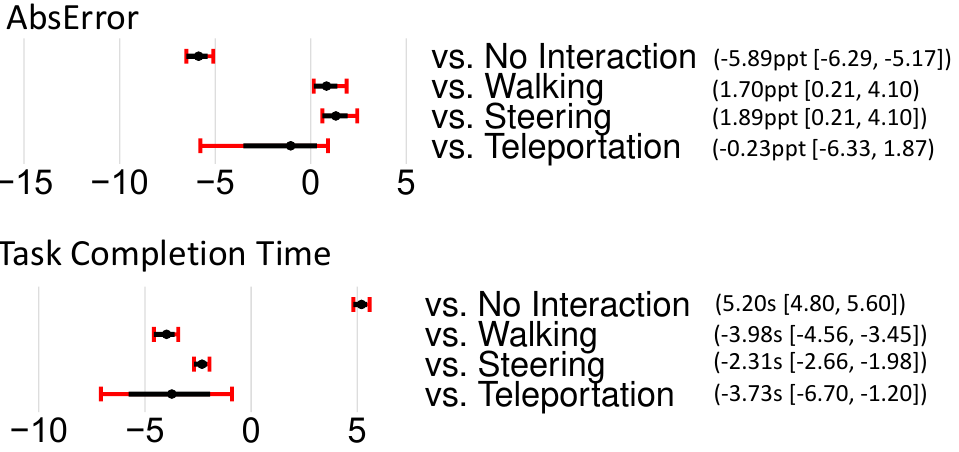}
\caption{Study 2 Results: pairwise comparison results between \studyFactors{Steering} with \studyFactors{Personal} and interaction techniques with \studyFactors{Frontal}}
\includegraphics[width=1\columnwidth]{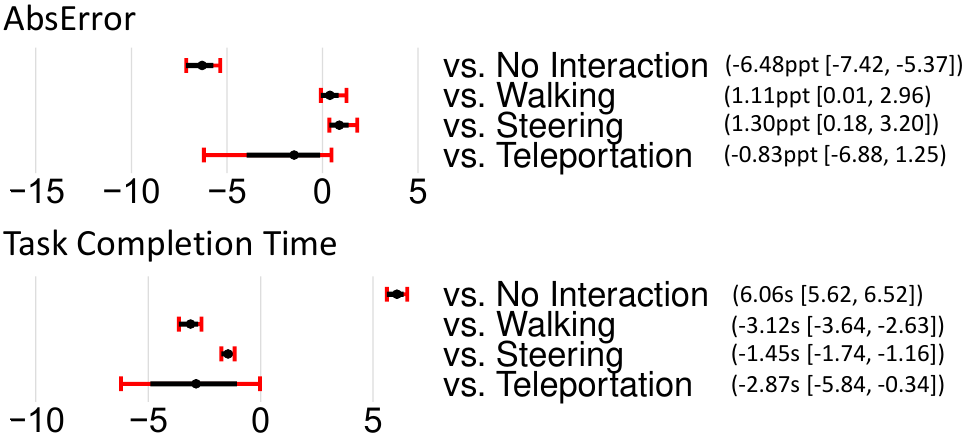}
\caption{Study 2 Results: pairwise comparison results between \studyFactors{Teleportation} with \studyFactors{Personal} and interaction techniques with \studyFactors{Frontal} }
\end{figure}








\end{document}